\documentclass[fleqn,usenatbib]{mnras}
\usepackage[T1]{fontenc}
\usepackage[varg]{newtx}
\usepackage{amsmath}
\usepackage{graphicx}
\usepackage{orcidlink}
\AtBeginDocument{}
\usepackage{xspace}
\newcommand\Msun{\ensuremath{\mathrm{M}_{\odot}}\xspace}
\newcommand{\tablefoot}[1]{\par\smallskip\noindent\parbox{\hsize}{\footnotesize \textbf{Notes.} #1}}

\title[Early AT2017gfo spectra from scaled 3D ejecta]{Early spectra of AT2017gfo from scaled 3D merger ejecta: mass--velocity distribution, 3D structure, and treatment of line opacity}

\author[L. J. Shingles et al.]{Luke J. Shingles,$^{1,2}$\orcidlink{0000-0002-5738-1612}\thanks{E-mail: luke.shingles@gmail.com}
Christine E. Collins,$^{3}$\orcidlink{0000-0002-0313-7817}
Vimal Vijayan,$^{2,4}$\orcidlink{0000-0002-4690-2515}
Andreas Fl\"{o}rs,$^{2}$\orcidlink{0000-0003-2024-2819}
Oliver Just,$^{2}$\orcidlink{0000-0002-3126-9913}
\newauthor
Fiona McNeill,$^{1}$\orcidlink{0009-0001-9528-7475}
Gerrit Leck,$^{2,5}$\orcidlink{0000-0002-0093-0211}
Andreas Bauswein,$^{2}$\orcidlink{0000-0001-6798-3572}
Gabriel Mart{\'i}nez-Pinedo,$^{2,5}$\orcidlink{0000-0002-3825-0131},
\newauthor
and Stuart A. Sim,$^{1}$\orcidlink{0000-0002-9774-1192}
\\
$^{1}$Astrophysics Research Centre, School of Mathematics and Physics, Queen's University Belfast, Belfast BT7 1NN, UK\\
$^{2}$GSI Helmholtzzentrum f\"ur Schwerionenforschung, Planckstra{\ss}e 1, 64291 Darmstadt, Germany\\
$^{3}$School of Physics, Trinity College Dublin, The University of Dublin, Dublin 2, Ireland\\
$^{4}$Department of Physics, University of Jyväskylä, P.O. Box 35, 40014 Jyväskylä, Finland\\
$^{5}$Institut für Kernphysik (Theoriezentrum), Fachbereich Physik,
Technische Universität Darmstadt, Schlossgartenstraße 2, 64289 Darmstadt, Germany}
\date{Accepted XXX. Received YYY; in original form ZZZ}
\pubyear{2026}

\begin{document}
\label{firstpage}
\pagerange{\pageref{firstpage}--\pageref{lastpage}}
\maketitle

\begin{abstract}
We present ARTIS radiative transfer calculations for neutron star merger ejecta with line-by-line opacities, wavelength-calibrated lanthanide data, and time-dependent r-process compositions and thermalisation, testing how the early kilonova depends on the mass--velocity distribution, dimensionality, and line treatment.
We rescale 3D dynamical ejecta from a merger simulation in density and velocity to approximate AT2017gfo.
Mass scaling alone provides good agreement with the bolometric light curve but inconsistent spectra.
With velocities additionally halved, the model also reproduces the observed spectra at 1.4, 2.4, and (albeit less accurately) 3.4 days.
The \ion{Sr}{ii} triplet contributes most to the 10\,500~\AA\ peak, and \ion{La}{iii} and \ion{Gd}{iii} to the 13\,500~\AA\ shoulder.
Although hydrodynamically not fully self-consistent, this model is to our knowledge the first kilonova model based on calibrated atomic data that reproduces the light curve and main spectral features of the early AT2017gfo, suggesting that total (dynamical plus post-merger) ejecta with higher mass, lower velocity, and similar composition compared to our dynamical ejecta could explain AT2017gfo.
A 2D azimuthal average mixes lanthanide-poor with lanthanide-rich regions, smooths the density, and makes the averaged model three times fainter at 1 day and redder throughout than the 3D model, highlighting the importance of 3D modelling.
We compare the line-by-line macroatom treatment with expansion-opacity and line-binned schemes with thermal redistribution, using identical atomic data.
Luminosities agree to $\lesssim$10~per cent after 1 day, but the expansion-opacity scheme, unlike converged line-binned schemes, overestimates the spectral flux by up to 20~per cent, which can bias inferred abundances.
\end{abstract}


\section{Introduction}
\label{sec:intro}

A kilonova is a rapidly evolving transient that follows a compact object merger and is powered by the radioactive decay of newly synthesised heavy nuclei \citep[e.g.,][]{metzger_electromagnetic_2010,metzger_kilonovae_2019}, produced through rapid neutron capture (the r-process).
To date, the most complete observed sequence of spectra of a kilonova is that of AT2017gfo \citep{coulter_swope_2017,smartt_kilonova_2017,pian_spectroscopic_2017}, the electromagnetic transient associated with the gravitational-wave signal GW170817 \citep{abbott_gw170817_2017}, which originated from a neutron star merger \citep[see][for a review]{margutti_first_2021}.
These spectra carry signatures of freshly synthesised r-process elements \citep[e.g.,][]{pian_spectroscopic_2017,smartt_kilonova_2017,watson_identification_2019}, which merger ejecta are expected to produce \citep[e.g.,][]{lattimer_blackholeneutronstar_1974,lattimer_decompression_1977,freiburghaus_rprocess_1999,goriely_rprocess_2011,korobkin_astrophysical_2012,fernandez_delayed_2013,wanajo_production_2014,perego_neutrinodriven_2014,just_comprehensive_2015,radice_binary_2018}.

An outstanding goal of recent kilonova modelling efforts is to reproduce the observations from AT2017gfo accurately enough to infer, ideally in an unambiguous manner, the multidimensional mass--velocity distribution and composition of the ejecta.
While individual properties of AT2017gfo can be reproduced using models that are simplified in one way or another, models that agree in both the light curve and spectral features (and their time evolution) pose an enormous multi-disciplinary challenge, because they require accurate knowledge of atomic data for the photon opacities as well as nuclear data for the composition and radioactive heating of the ejecta, and computationally expensive numerical hydrodynamics and radiative transfer simulations that link ejecta properties to observables.
Each of these modelling ingredients is still beset with significant uncertainties.
A particularly important aspect for spectral modelling is to calibrate theoretically obtained atomic lines to experimentally known values, which is necessary for a faithful association of observed spectral features with specific ions \citep[e.g.,][]{ryabchikova_major_2015,kurucz_including_2018,domoto_signatures_2021,domoto_lanthanide_2022,gillanders_modelling_2022,flors_opacities_2023,flors_calibrated_2026}.
So far no fully self-consistent 3D hydro-plus-kilonova model using calibrated atomic data has been shown to reproduce both the light curve and spectral features of AT2017gfo during its early photospheric phase.

Kilonova calculations that treat thermalisation of radioactive decay energy in detail \citep[e.g.,][]{hotokezaka_radioactive_2020,andalman_betaparticle_2026} or use wavelength-calibrated line lists \citep[e.g.,][]{gillanders_modelling_2022} have mostly used simplified ejecta models, such as 1D density profiles with parameterised compositions.
Multidimensional radiative transfer studies have likewise often adopted parameterised or spatially averaged ejecta distributions \citep[e.g.,][]{wollaeger_impact_2018,kawaguchi_radiative_2018,bulla_possis_2019,wollaeger_broad_2021}.
Such averaging removes any small-scale structure of the merger ejecta, such as angular variations of the composition, dense clumps, and low-density channels between them.
3D radiative transfer calculations that keep this structure have been presented for the ejecta of neutron star mergers \citep{tanaka_radiative_2013,collins_3d_2023,shingles_selfconsistent_2023,neuweiler_longterm_2023,collins_exploring_2026,kiuchi_firstprinciples_2026} and of black hole--neutron star mergers \citep{tanaka_radioactively_2014,darbha_electromagnetic_2021,kawaguchi_three_2024}.
Such calculations indicate that the 3D structure can be important for predicting synthetic observables. 
For example, in \citet{shingles_selfconsistent_2023}, we showed that a sequence of spectra similar to that of AT2017gfo can be obtained from a model that includes the 3D merger structure while a model from a 1D spherical average had strongly suppressed flux below $\sim$8000~\AA ~(and different luminosity), which is less consistent with the observed spectral series.
However, the model used by \citet{shingles_selfconsistent_2023} was too faint and evolved too rapidly, probably because its mass was low.
It was based on a simulation that ended 20~ms after the merger and thus contained only the dynamical ejecta (matter that tidal forces and shocks expel during the first milliseconds of the merger) with no post-merger ejecta.
As such, its mass of 0.005~\Msun is well below the total ejecta mass estimated for AT2017gfo, which is 0.03--0.05~\Msun as inferred from its light curve \citep{smartt_kilonova_2017}, and up to $\sim$0.08~\Msun as suggested by multicomponent fits \citep{kasen_origin_2017,villar_combined_2017}.
The spectral features of that model were also at shorter wavelengths than those of AT2017gfo, which suggested that the expansion velocities of the model were too high.
Of the 3D calculations listed above, \citet{kawaguchi_three_2024} and \citet{kiuchi_firstprinciples_2026} include the post-merger ejecta, but they present no synthetic spectra and use wavelength-calibrated data for only eight ions.

Kilonova models also differ in how they treat the line opacity.
A line-by-line treatment evaluates the optical depth of each line individually in the Sobolev approximation \citep{sobolev_moving_1960} and follows the emission after each absorption, for example with the macroatom method of \citet{lucy_monte_2002,lucy_monte_2003}.
To reduce the computational cost, many kilonova calculations instead combine the lines of each wavelength bin into one binned opacity, either an expansion opacity \citep{karp_opacity_1977,eastman_spectrum_1993} or a line-binned opacity \citep{fontes_linebinned_2020}.
Codes that use binned opacities typically also assume thermal redistribution, in which the re-emitted radiation has a spectrum set by the local temperature and opacity of the gas, with no memory of the transition that absorbed it \citep[e.g.,][]{wollaeger_impact_2018,bulla_possis_2019}.

This work has three aims.
First, we want to understand whether simple modifications to our previously adopted 3D dynamical-ejecta model can lead to luminosities and timescales that match the light curve of AT2017gfo while also approximately reproducing the main features of the spectra.
Ideally, an exploration of more luminous models would be based on long-term hydrodynamic simulations that capture the post-merger ejecta, such as viscously driven black-hole torus outflows \citep{fernandez_delayed_2013,siegel_threedimensional_2018,fujibayashi_viscous_2020}, neutrino-driven winds \citep[e.g.,][]{perego_neutrinodriven_2014,fujibayashi_postmerger_2020,just_endtoend_2023}, spiral-wave winds \citep{nedora_spiralwave_2019}, or magnetically driven ejecta \citep[e.g.,][]{combi_jets_2023,curtis_magnetized_2024,kiuchi_firstprinciples_2026}. 
However, since 3D modelling of the long-term post-merger evolution is computationally very expensive, in place of such a simulation here we rescale the mass and velocities of the 3D dynamical ejecta.
Although this ejecta model is not fully self-consistent, the purpose is to explore which ejecta properties (i.e., mass--velocity distribution and composition) are sufficient for the resulting kilonova light curve and spectra to resemble AT2017gfo.
Knowledge of these properties would provide constraints on the combined (dynamical and post-merger) mass distribution and composition and represent a useful point of reference for future kilonova models involving more self-consistent post-merger ejecta.

Second, motivated by the current availability of families of 2D models \citep[e.g.,][]{kawaguchi_lowmass_2021, just_endtoend_2023, sippensgroenewegen_2d_2025}, we want to understand how the 3D structure of merger ejecta changes the light curves and spectra relative to 2D-averaged ejecta.

Third, we want to quantify how much the common approximations for the line opacity and for the emission that follows a line absorption, the expansion opacity and thermal redistribution (see Sections~\ref{sec:methodsopac} and \ref{sec:methodsemission} for definitions and discussion), change the synthetic spectra and thus could impact how well models can be used to infer ejecta parameters from observations.
This is particularly relevant to future studies of multidimensional models in which a trade-off between high precision and computational efficiency may be needed.

To investigate these aims, we present 3D radiative transfer calculations that combine wavelength-calibrated lanthanide atomic data \citep{flors_calibrated_2026} with these scaled models.

Section~\ref{sec:methods} describes the methods and models, Section~\ref{sec:resultsscaled} compares the scaled models with AT2017gfo and presents the ion and transition contributions, and Section~\ref{sec:resultsopac} quantifies the effect of the binned-opacity and thermal-redistribution approximations.
Section~\ref{sec:discussion} discusses the implications and the limitations of these results, and Section~\ref{sec:conclusions} gives our conclusions.

\section{Methods and models}
\label{sec:methods}
\subsection{Base ejecta model, nuclear heating, and radiative transfer}
\label{sec:rt}
We use the multidimensional Monte Carlo radiative transfer code ARTIS \citep{sim_multidimensional_2007,kromer_timedependent_2009}, which has been extended to handle kilonova ejecta \citep{collins_3d_2023,shingles_selfconsistent_2023}.
ARTIS represents the radiation field and the decay energy as indivisible Monte Carlo energy packets \citep{lucy_computing_1999,lucy_monte_2002}, each of which is followed through the ejecta.

Our base 3D ejecta model is the same model as in \citet{collins_3d_2023} and \citet{shingles_selfconsistent_2023}.
The ejecta originate from a 3D relativistic smoothed-particle hydrodynamics (SPH) simulation \citep{oechslin_conformally_2002,oechslin_relativistic_2007,bauswein_testing_2010} of the merger of two 1.35~\Msun neutron stars with the SFHo equation of state \citep{steiner_corecollapse_2013}.
In this simulation, the neutrinos are treated with the improved leakage-equilibration-absorption scheme of \citet{ardevol-pulpillo_improved_2019}, and the system is followed up to 20~ms after the merger.
A nuclear reaction network is then run in post-processing on the thermodynamic trajectory of each SPH particle \citep[see][for details]{collins_3d_2023}, with the densities beyond 20~ms extrapolated under the assumption of homologous (ballistic) expansion, in which each particle moves radially at a constant velocity.

At 0.1 days, we map the SPH particles and their network results onto a 3D grid to produce a snapshot of the composition, density, and thermal energy.
The mapping is that of \citet{shingles_selfconsistent_2023}, and the velocity field of the grid is purely radial.
The thermal energy of each particle is the time integral of the nuclear energy release rate from the network up to the snapshot time $t_\mathrm{snap}$, with each contribution reduced by the adiabatic factor $t/t_\mathrm{snap}$, where $t$ is the time of the release.
This estimate assumes that the released energy is trapped locally, and it includes the neutrino energy of the $\beta$ decays, which does not heat the gas, so it overestimates the trapped energy.
ARTIS releases this energy as radiation packets at the start of the first timestep.
In the model with halved velocities (Section~\ref{sec:ejectamodels}), the snapshot energy supplies about half of the escaping luminosity at 0.3 days, about a quarter at 1 day, about a fifth at 1.4 to 2.4 days, and less than a tenth after 5 days.
If the snapshot energy is too high by a factor of two, the model luminosity is thus too high by up to $\sim$10~per cent at 1.4 to 2.4 days, and by more at earlier times.

ARTIS follows the $\alpha$, $\beta$, and fission decays in each cell with a set of Bateman equations and tabulated nuclear data, which include mean lifetimes, branching ratios, $\gamma$-ray spectra, and the average energy of emitted $\alpha$ and $\beta$ particles.
The only fission channel is the spontaneous fission of $^{254}$Cf, which \citet{zhu_californium254_2018} identified as the dominant fission contribution to the late-time heating of kilonovae.
Its decay data come from ENDF/B-VII.1 \citep{chadwick_endf_2011} and its fission yields from the GEF model \citep{schmidt_general_2016}.
This channel is an addition to the treatment of \citet{shingles_selfconsistent_2023}, which neglected fission, and its effect on the total energy release rate is small for this ejecta model.
The elemental abundances in each cell change with time during the radiative transfer, whereas many kilonova models hold the composition fixed in time \citep[e.g.,][]{kasen_origin_2017,wollaeger_broad_2021}.

The thermalisation efficiency of the decay products, the fraction of their energy that the gas absorbs, falls with time and strongly affects kilonova luminosities \citep{barnes_radioactivity_2016,hotokezaka_radioactive_2020,barnes_kilonovae_2021}.
The $\gamma$-rays are transported with frequency-dependent opacities for Compton scattering, photoelectric absorption, and pair production.
The $\alpha$ and $\beta$ particles deposit their energy at the location of the decay, but not instantly.
As in \citet{shingles_selfconsistent_2023}, ARTIS follows the kinetic energy of each particle packet in time.
The packet loses energy to the gas at the local density, with the density-scaled average energy-loss rate of \citet[][their figure 6]{barnes_radioactivity_2016}, so the thermalisation lags behind the decays.
We improve on the treatment of \citet{shingles_selfconsistent_2023} by including the adiabatic loss of the particle energy due to expansion, although this loss changes the deposited energy only a little.
The thermalised fraction follows from the density history of each packet and the initial kinetic energy of the decay that produced the packet, not from an analytic approximation for the thermalisation efficiency \citep[see, e.g.,][]{barnes_radioactivity_2016}.
The kinetic energy of the fission fragments is deposited instantly at the location of the decay.

All calculations assume local thermodynamic equilibrium (LTE) at every timestep.
A single electron temperature $T_e$ per cell follows from the Monte Carlo estimator of the mean intensity with the Stefan--Boltzmann law, and the Saha and Boltzmann equations at $T_e$ give the ionisation and excitation.

The radiative transfer uses 64 logarithmically spaced timesteps from 0.1 to 80 days.
Light-travel-time effects limit the range of times in which the synthetic observables are complete \citep{shingles_selfconsistent_2023}: from $\sim$0.2 to 14 days for the models with the original velocities, and from $\sim$0.1 to 47 days for the models with halved velocities (Section~\ref{sec:ejectamodels}).
Each calculation uses of the order of $10^8$ Monte Carlo packets.\footnote{The line-by-line macroatom calculation for the model with halved velocities (Section~\ref{sec:ejectamodels}) used $4\times10^8$ packets and needed $9\times10^5$ core hours, or 18 days of wall time on 2112 cores (12 nodes with AMD Zen~4 processors) of the Virgo cluster at GSI.}

All synthetic light curves and spectra in this work are for the polar direction of \citet{shingles_selfconsistent_2023}, which we define below.
The $z$-axis is perpendicular to the orbital plane of the merger.
The polar direction is the viewing-angle bin within $37^\circ$ of the negative $z$-axis ($180^\circ-\theta \leq 37^\circ$, where $\theta$ is the angle from the positive $z$-axis as defined by \citealt{collins_3d_2023}), spanning all azimuthal angles.
The luminosities are isotropic-equivalent values: the energy that escapes into this bin, scaled by the ratio of $4\pi$ to the solid angle of the bin.
The fluxes at 40~Mpc, the distance that we adopt for AT2017gfo (Section~\ref{sec:obsdata}), follow from the same scaled energy and are what an observer in this direction would measure.
This bin contains the observer inclinations inferred for AT2017gfo from very long baseline interferometry imaging and afterglow modelling of the associated jet \citep[$\sim$15$^\circ$;][]{ghirlanda_compact_2019} and from the superluminal motion of the jet \citep[19--25$^\circ$;][]{mooley_optical_2022}.
\citet{collins_3d_2023} and \citet{shingles_selfconsistent_2023} present the dependence on the viewing angle for the unscaled model.

\subsection{Atomic data}
\label{sec:atomicdata}
The National Institute of Standards and Technology (NIST) Atomic Spectra Database \citep{kramida_nist_2024} contains measured energy levels and transitions, but these data cover only a small part of the transitions that contribute to the opacity, and for many ions even the measured levels are incomplete.
Theoretical calculations, such as those in the Japan-Lithuania database \citep{tanaka_systematic_2020,kato_japanlithuania_2021}, are more complete, but their wavelengths have large uncertainties.
Measured level energies place the transitions at the correct wavelengths, and completeness gives the correct overall opacity.
For example, \citet{shingles_selfconsistent_2023} show that calibrated Sr, Y, and Zr data place the prominent \ion{Sr}{ii} feature at the correct wavelength, whereas uncalibrated data put the \ion{Sr}{ii} triplet at 18\,000--20\,000~\AA\ instead of 10\,000--11\,000~\AA.
Completeness matters as well, as \citet{gillanders_improved_2026} find a lanthanide mass fraction in the line-forming region of AT2017gfo at 3.4 days that is 20 times lower than an earlier estimate with less complete lanthanide line lists \citep{gillanders_modelling_2022}.
\citet{flors_calibrated_2026} replace the computed energy of each level with the measured value (mainly from the NIST database) where one exists, which corrects the wavelengths of the transitions between such levels.
Their line lists are thus both extensive and more accurate in wavelength.
Since the calibration does not change the computed structure of the levels, the line strengths stay theoretical, and the transition probabilities change only through their dependence on the transition energy.

Our starting point is the AD2 dataset of \citet{shingles_selfconsistent_2023}.
AD2 uses the CMFGEN compilation \citep{hillier_iterative_1990,hillier_treatment_1998} for selected elements with atomic number $Z\leq 28$ and the Japan-Lithuania database \citep[version 1.1;][]{tanaka_systematic_2020,kato_japanlithuania_2021} for $Z = 29$--$88$, except for Sr, Y, and Zr, which use calibrated data from \citet{kurucz_including_2018}.
The transition probabilities of these Sr, Y, and Zr data include the correction from the erratum to \citet{shingles_selfconsistent_2023} \citep{shingles_erratum_2025}.
In our new dataset, `AD3', we replace the level and transition data for the singly and doubly ionised stages of elements $Z = 57$--$70$ with wavelength-calibrated data by \citet{flors_calibrated_2026}.
We also update to version 2.1 of the Japan-Lithuania database, which includes calculations for additional ions by \citet{gaigalas_extended_2019} and \citet{radziute_extended_2020}.\footnote{\url{http://dpc.nifs.ac.jp/DB/Opacity-Database/}}
AD3 also contains the \ion{Th}{iii} data of \citet{kitoviene_theoretical_2025}.

In total, our AD3 atomic dataset contains about 49 million bound--bound transitions across 323 ionisation stages of 83 elements (about four stages for each element).
For 68 of the 83 elements, these stages are the neutral atom and the first three ions, while seven elements extend to the fourth ion and the remaining eight elements, such as H, He, and Th, have fewer stages.
AD3 contains no photoionisation cross-sections, so our calculations include no bound--free opacity.
\subsection{Ejecta model variations}
\label{sec:ejectamodels}

\begin{table}
\caption{Summary of the ejecta models compared in this work.
All models use the AD3 atomic dataset.
The grid gives the number of cells along $x$, $y$, and $z$ for the 3D models, and along the cylindrical radius and $z$ for the 2D models.
The extent $v_\mathrm{max}$, in units of the speed of light $c$, is the half-width of the Cartesian grid along each axis, and for the 2D models the maximum cylindrical radius and the half-height of the cylindrical grid.
The cells are uniform, so the cell width is $2v_\mathrm{max}$ divided by the number of cells along $x$, $y$, or $z$, and $v_\mathrm{max}$ divided by the number of cells along the cylindrical radius.
The $v$-scale is the factor applied to the radial velocity of every cell at a fixed total mass, and $E_\mathrm{kin}$ is the total non-relativistic kinetic energy of the ejecta, from an integral over the volume of each cell and a sum over the cells.}
\label{tab:modelsummary}
\centering
\setlength{\tabcolsep}{3.5pt}
\begin{tabular}{lccccrr}
\hline\hline
Model & Dim. & Grid & $v_\mathrm{max}$ & Mass & $v$-scale & $E_\mathrm{kin}$\\
 & & & ($c$) & (\Msun) & & ($10^{50}$ erg)\\
\hline
dyn (unscaled)         & 3D & $50^3$ & 0.48 & 0.005 & $1.0$ & 2.51\\
dyn 8x mass            & 3D & $50^3$ & 0.48 & 0.04 & $1.0$ & 20.0\\
dyn 8x mass v/2        & 3D & $50^3$ & 0.24 & 0.04 & $0.5$ & 5.01\\
dyn 8x mass v/2        & 2D & $25{\times}50$ & 0.24 & 0.04 & $0.5$ & 4.45\\
dyn 8x mass v/2        & 2D & $50{\times}100$ & 0.24 & 0.04 & $0.5$ & 4.36\\
\hline
\end{tabular}
\end{table}

\begin{figure}
\centering
\includegraphics[width=\hsize]{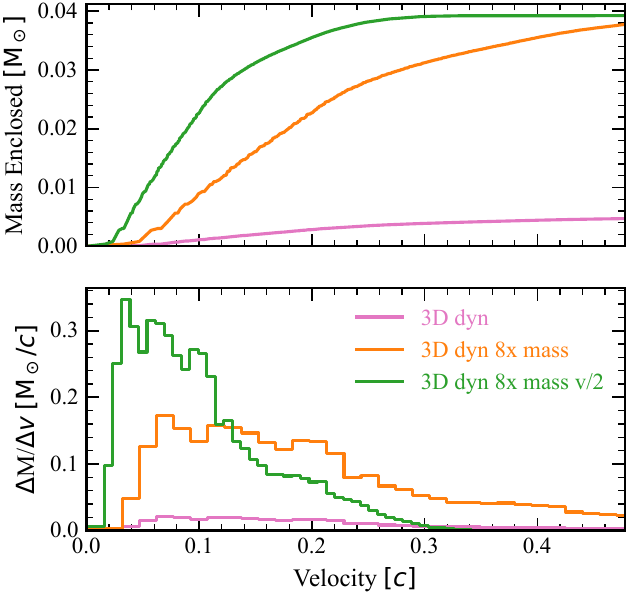}
   \caption{Radial ejecta profiles for the unscaled and scaled 3D models: cumulative enclosed mass (top) and mass per radial-velocity interval $\Delta M / \Delta v$ (bottom) as functions of the radial velocity $v_r$.
The bin width of $\Delta M / \Delta v$ is the cell width of each model: $0.019c$ for the unscaled and mass-scaled models and $0.0095c$ for the velocity-compressed model.
The horizontal axis ends at $0.48c$, the half-width of the Cartesian grid of the unscaled and mass-scaled models.
The edge regions of the grid beyond this radius extend to $0.83c$ and hold 4~per cent of the mass of these models.
The azimuthally averaged 2D models share the radial profile of their 3D parent model (dyn 8x mass v/2) below $0.24c$ by construction.}
     \label{fig:density}
\end{figure}

Table~\ref{tab:modelsummary} summarises the five ejecta models compared in this work.
We apply scaling factors to the density and expansion velocity of the original model on the $50^3$-cell grid of \citet{shingles_selfconsistent_2023}.
These scalings test how the mass--velocity distribution affects the sequence of spectra, and in particular whether the agreement with the brighter and more slowly evolving AT2017gfo improves.
The scaled models are not self-consistent merger simulations, but they let us vary the mass and the velocities while keeping the 3D structure of a merger simulation.
In `dyn 8x mass' (hereafter the mass-scaled model), we multiply the density in every cell by eight, which leaves the velocity structure unchanged.
We chose this factor because it raises the mass to 0.04~\Msun, within the 0.03--0.05~\Msun that \citet{smartt_kilonova_2017} estimated for AT2017gfo from its light curve.
In `dyn 8x mass v/2' (hereafter the velocity-compressed model), we keep the mass at 0.04~\Msun and also divide all expansion velocities by two, which makes the ejecta slower and denser.
The velocity scaling halves the radial velocity of every cell and keeps the number of cells, so the grid extends to $0.24c$ instead of $0.48c$, where $c$ is the speed of light, and the cells are half as wide (Table~\ref{tab:modelsummary}).
We reduce the velocities rather than raise them, because the spectral features of the unscaled model lie blueward of the observed features (Section~\ref{sec:intro}).
This blueward offset indicates that its expansion velocities are too high.
The mass scaling raises the kinetic energy by a factor of eight, and the velocity scaling then lowers it by a factor of four (Table~\ref{tab:modelsummary}).
Halving the velocities at fixed mass raises the density by a further factor of eight because the volume of each fluid element scales as $v_r^3$, where $v_r$ is the radial velocity.

Since the scalings keep the original elemental mass fractions, all models share the mass-weighted electron fraction $\langle Y_e\rangle=0.31$ and lanthanide mass fraction $X_\mathrm{lan}=0.023$ (summed from La to Lu, at 0.1 days) of the dynamical ejecta.
Lanthanide mass fractions of only $\sim$$10^{-3}$ already raise the opacity by orders of magnitude \citep{kasen_opacities_2013}, and \citet{kasen_origin_2017} use $X_\mathrm{lan}\gtrsim10^{-2}$ for `red' (lanthanide-rich) ejecta and $X_\mathrm{lan}\sim10^{-4}$ for `blue' (lanthanide-poor) ejecta.
With $X_\mathrm{lan}=0.023$, our models are thus lanthanide-rich on average.
We call matter with $X_\mathrm{lan} < 10^{-3}$ lanthanide-poor.

The composition varies strongly with latitude within the ejecta.
Within $37^\circ$ of the negative $z$-axis, the mass-weighted electron fraction is 0.43 and the lanthanide mass fraction is $\sim$0.004, and regions with $X_\mathrm{lan} < 10^{-3}$ hold about 70~per cent of this polar mass.
This polar matter is 6~per cent of the ejecta mass, whereas the equatorial region within $37^\circ$ of the $x$--$y$ plane holds about 70~per cent of the mass and has $X_\mathrm{lan}\approx0.03$.
The composition also varies with azimuth at a given latitude, so the lanthanide-poor and the lanthanide-rich matter are interleaved in three dimensions.
In the ejecta as a whole, regions with $X_\mathrm{lan} < 10^{-3}$ hold about a quarter of the mass, which is $\sim$0.01~\Msun in the scaled models.
More than half of this matter has $X_\mathrm{lan} < 10^{-4}$, and regions with $X_\mathrm{lan} < 10^{-2}$ hold about half of the total mass.
The mean lanthanide mass fraction thus describes a mixture of lanthanide-rich and lanthanide-poor matter and not a uniform composition.

The lanthanide mass fraction also varies with radius, rising in the velocity-compressed model from $\sim$0.014 inside $0.05c$ to $\sim$0.03 between 0.1 and $0.25c$, so the fast outer regions are richer in lanthanides than the average.
The values of $\langle Y_e\rangle$ and $X_\mathrm{lan}$ depend on the neutrino treatment of the merger simulation \citep{ardevol-pulpillo_improved_2019} and on the nuclear-physics inputs of the network, such as the nuclear mass model and the fission treatment \citep[e.g.,][]{barnes_kilonovae_2021}.

With these scalings, all of the mass has the composition of the dynamical ejecta.
Post-merger ejecta are expected to be poorer in lanthanides, and \citet[][their table 1]{just_endtoend_2023} give a combined mass fraction of lanthanides and actinides below 0.01 for the post-merger ejecta of their end-to-end models.
For the total ejecta of these models, the values of the same table imply a combined mass fraction of lanthanides and actinides of 0.01--0.03, compared with 0.04 in our models.
The distribution of these elements also differs, as the end-to-end models have most of them in the dynamical ejecta, whereas those of our scaled models are spread over the whole mass.

Figure~\ref{fig:density} shows the radial profiles of the three 3D models.
Mass scaling alone preserves the shape of the velocity distribution, whereas the velocity-compressed model concentrates its mass at low velocities.
Half of its mass lies below $v_r \sim 0.09c$, the mass distribution $\Delta M / \Delta v$ peaks near $0.04c$ at twice the peak value of the mass-scaled model, and only 0.4~per cent of the mass lies above $0.3c$, in the edge regions of its grid.

To explore the effect of 3D asymmetries on the light curves and spectra, we also construct a 2D axisymmetric version of the velocity-compressed model, the model that best reproduces the early sequence of spectra.
We average the 3D density, composition, and initial thermal energy azimuthally onto uniform cylindrical grids of $25{\times}50$ and $50{\times}100$ cells (cylindrical radius by $z$), which extend to $0.24c$ in cylindrical radius and from $-0.24c$ to $0.24c$ in $z$.
The cylindrical grids do not cover the edge regions of the Cartesian grid, where the cylindrical radius is larger than half the width of the grid, and the 2D models therefore have 2--3~per cent less mass than their 3D parent.
This missing mass is at the highest velocities, so the 2D models have about 10~per cent less kinetic energy (Table~\ref{tab:modelsummary}).

\begin{figure}
\centering
\includegraphics[width=\hsize]{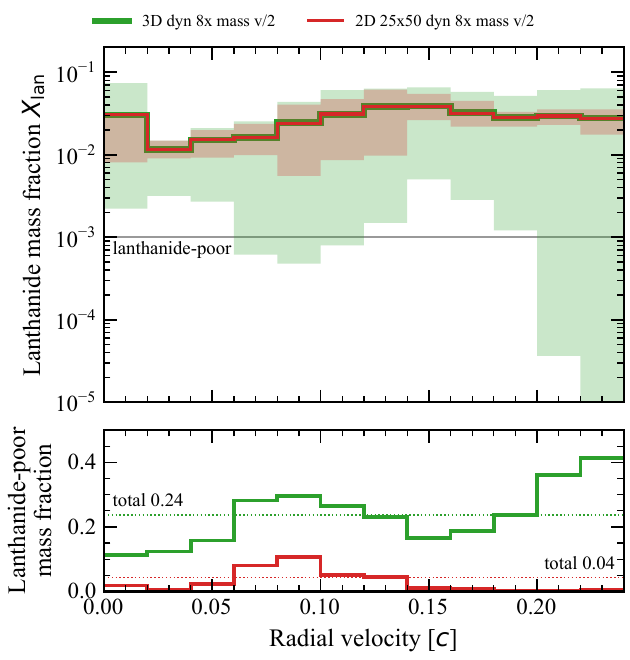}
   \caption{Lanthanide mass fraction $X_\mathrm{lan}$ as a function of the radial velocity for the velocity-compressed 3D model and its 2D azimuthal average on the $25{\times}50$ grid.
In the top panel, the line of each model is the mass-weighted mean of $X_\mathrm{lan}$ in bins of width $0.02c$ in the radial velocity of the cells, and the two lines almost coincide.
The shaded band of a model contains the central half of the mass of each bin, from the 25th to the 75th percentile in $X_\mathrm{lan}$.
The lower edge of the band of the 3D model lies below the range of the axis in the outermost bin.
Matter below the horizontal line at $X_\mathrm{lan} = 10^{-3}$ is lanthanide-poor, and a band extends below this line in the bins in which more than a quarter of the mass is lanthanide-poor.
The bottom panel gives the fraction of the mass of each bin that is in lanthanide-poor cells.
The dotted lines give this fraction for all cells of each model, which include the cells beyond the velocity range of the figure.}
     \label{fig:xlanvelocity}
\end{figure}

The azimuthal average mixes the composition (Fig.~\ref{fig:xlanvelocity}).
The average leaves the mass-weighted mean of $X_\mathrm{lan}$ at each radial velocity almost unchanged, but it narrows the distribution of $X_\mathrm{lan}$ around this mean, mainly by removing the matter with the lowest $X_\mathrm{lan}$.
Regions with $X_\mathrm{lan} < 10^{-3}$ hold only a few per cent of the mass of the 2D models, compared with about a quarter in the 3D model, in which this lanthanide-poor matter is present at all radial velocities.
The average also smooths the density, because it conserves the mass at each cylindrical radius and height but removes the density contrast in azimuth.
The mass-weighted mean density of the 2D models is therefore lower than that of the 3D model, by a factor of $\sim$1.3 below $0.1c$ and by a factor of 2--3 above $0.15c$.

Unless we state otherwise, `the 2D model' is the model on the $25{\times}50$ grid, which has the same cell width as the 3D grid.
The $50{\times}100$ grid halves this width, and we use it only as a control for the comparison with the 3D model, to test whether the differences come from the azimuthal averaging or from a sensitivity of the results to the number of cells in the grid (Section~\ref{sec:2dcomparison}).

\subsection{Binned opacities}
\label{sec:methodsopac}

By default, ARTIS uses detailed line-by-line opacities.
In this treatment, line optical depths are evaluated using the Sobolev approximation \citep{sobolev_moving_1960}, and a packet interacts with line $i$ only at the Sobolev resonance point, i.e., where the comoving wavelength of the packet is equal to the line rest wavelength $\lambda_i$.
The packet interacts there with probability $1-e^{-\tau_i}$, where $\tau_i$ is the Sobolev optical depth of the line.
However, many published kilonova calculations reduce the computational cost by grouping lines into wavelength bins, each with one opacity value.
Such binned opacities can be used in multigroup schemes and in diffusion schemes for optically thick regions \citep[e.g.,][]{wollaeger_radiation_2013,fontes_linebinned_2020}.

For this paper, we have implemented a binned opacity in ARTIS to test how sensitive the results are to the opacity treatment when the code, the ejecta model, and the atomic data are the same.
Section~\ref{sec:binnedopacdef} defines the two binned opacity formulations that we consider, the expansion opacity and the line-binned opacity.
To elucidate the distinction, Section~\ref{sec:binTransport} compares how each treatment transports a packet across one wavelength bin.
Section~\ref{sec:cappedopac} then introduces a cap that makes the line-binned opacity tractable in ARTIS, and Section~\ref{sec:binnedimpl} describes our implementation.

\subsubsection{Expansion and line-binned opacities}
\label{sec:binnedopacdef}

A binned opacity replaces the lines in a wavelength bin of width $\Delta\lambda$ with one opacity $\kappa$ that is constant across the bin.
In homologous expansion and to first order in $v/c$, where $v$ is the ejecta velocity, the comoving wavelength $\lambda$ of a packet increases by $\Delta\lambda$ over the path length
\begin{equation}
   \Delta s = \frac{ct\,\Delta\lambda}{\lambda},
   \label{eq:deltas}
\end{equation}
where $t$ is the time since the merger.
The optical depth across the bin is then
\begin{equation}
   \tau_\mathrm{bin} = \kappa\rho\,\Delta s,
   \label{eq:taubin}
\end{equation}
where $\rho$ is the local mass density.
Each binned opacity follows from a choice of the optical depth that each line adds to $\tau_\mathrm{bin}$.
The \emph{expansion opacity} \citep{karp_opacity_1977}, as formulated by \citet{eastman_spectrum_1993} and as used by, e.g., \citet{tanaka_systematic_2020} and \citet{bulla_critical_2023}, has each line add its Sobolev interaction probability, $1-e^{-\tau_i}$.
Line $i$ thus contributes the opacity $(1-e^{-\tau_i})/(\rho\,\Delta s_i)$, where $\Delta s_i = ct\,\Delta\lambda/\lambda_i$ is the path length over which a packet Doppler shifts across the wavelength bin, as given by equation~(\ref{eq:deltas}) evaluated at $\lambda = \lambda_i$.
The sum over the lines in the bin is
\begin{align}
    \kappa_{\mathrm{exp}} &= \frac{1}{ct\rho} \sum_i \frac{\lambda_i}{\Delta\lambda} \left(1-e^{-\tau_i}\right).
    \label{eq:expopac}
\end{align}
Within a narrow bin, $\lambda \approx \lambda_i$, so each line adds the optical depth $1-e^{-\tau_i}$ for every packet that crosses the bin.
The \emph{line-binned opacity} of \citet{fontes_linebinned_2020} instead lets each line add its full Sobolev optical depth, so this opacity has $\tau_i$ in place of $1-e^{-\tau_i}$ in equation~(\ref{eq:expopac}), yielding
\begin{align}
    \kappa_{\mathrm{line\text{-}bin}} &= \frac{1}{ct\rho} \sum_i \frac{\lambda_i}{\Delta\lambda} \tau_i.
    \label{eq:linebinnedopac}
\end{align}
Section~\ref{sec:cappedopac} introduces a cap on the contribution of each line to this sum, so we call this opacity the \emph{uncapped line-binned opacity}.

The expansion opacity and the line-binned opacity have distinct properties, and each accurately captures a different aspect of what occurs in the full line-by-line treatment.
In particular, as discussed in Section~\ref{sec:binTransport}, the line-binned opacity recovers the same probability of no interaction as the line-by-line treatment for pure line absorption of a packet as it Doppler shifts through a wavelength bin.
However, when applied to line scattering, the line-binned opacity can lead to significantly more interaction events than the line-by-line treatment at high optical depths.
\citet{fontes_linebinned_2020} argue that tracking the detailed diffusion of radiation may not be critical for accurately modelling kilonova emission.
An increased number of events can nevertheless become expensive for a Monte Carlo approach.
In contrast, the expansion opacity leads to a mean number of interactions more similar to that of the line-by-line treatment for a scattering-dominated case, but it does not recover the probability of no interaction for absorption (see Section~\ref{sec:binTransport} for examples of these differences).

Binned opacities have the advantage that they can be tabulated in advance, and, as discussed by \citet{fontes_linebinned_2020}, the line-binned opacity has the further benefit that its tabulation does not depend on time.
This follows because the Sobolev optical depth of a line is proportional to the population of its lower level, corrected for stimulated emission, and to $t$, since the velocity gradient in homologous expansion is $1/t$.
In LTE, the level populations follow from the local density, temperature, and composition.
Thus, the expansion opacity of a bin is a function of these local conditions and $t$.
However, in the line-binned opacity, the factor $t$ in each term $\tau_i$ cancels against the $1/(ct\rho)$ prefactor.
The line-binned opacity thus has no explicit time dependence, which significantly simplifies its tabulation for use in numerical simulations.
This opacity changes with time only through the level populations.

\subsubsection{Transport across one bin}
\label{sec:binTransport}

As noted by \citet{fontes_linebinned_2020}, the expansion opacity reduces to the line-binned result for $\tau_i \ll 1$.
Thus the difference between them is important only for strong lines.
To illustrate this difference and to compare with the line-by-line treatment, consider a packet that crosses one bin that contains several lines, where line $i$ has the Sobolev optical depth $\tau_i$.

First, we compare the probability $P_0$ that the packet crosses the bin without an interaction.
In the line-by-line treatment, the packet meets the lines one at a time at their resonance points, and it passes line $i$ without an interaction with probability $e^{-\tau_i}$.
The probability that it passes all the lines without an interaction is the product of these probabilities, $\exp(-\sum_i \tau_i)$.
With a binned opacity, the opacity is constant along the path of the packet in the bin, and the probability of no interaction is then
\begin{equation}
   P_0 = e^{-\tau_\mathrm{bin}}.
   \label{eq:ptr}
\end{equation}
For the uncapped line-binned opacity, $\tau_\mathrm{bin} = \sum_i \tau_i$ (from equations~\ref{eq:taubin} and \ref{eq:linebinnedopac}), so $P_0$ is equal to the line-by-line value.
However, the line-binned opacity does not locate the interactions with a strong line at the resonance point of the line.
A packet that enters a bin with $\tau_\mathrm{bin} \gg 1$ is absorbed within a small fraction of the bin, so these interactions are near the blue edge of the bin.
For such a packet, this wavelength error is at most $\Delta\lambda$, so it decreases as the bins become narrower, and it is consequential only if the scheme used is sensitive to the absorption wavelength.
For a packet that is emitted inside the bin, however, the distribution of the opacity within the bin sets the probability that the packet leaves the bin without a further interaction, which we discuss at the end of this section and in Section~\ref{sec:methodsemission}.

For the expansion opacity, $\tau_\mathrm{bin} = \sum_i (1-e^{-\tau_i})$ (from equations~\ref{eq:taubin} and \ref{eq:expopac}).
If all lines are weak ($\tau_i \ll 1$), this optical depth is approximately $\sum_i \tau_i$, so the expansion opacity also gives the line-by-line value of $P_0$.
However, for a bin with one strong line of optical depth $\tau_l \gg 1$, the line-by-line treatment gives $P_0 = e^{-\tau_l} \approx 0$, while the expansion opacity instead gives $\tau_\mathrm{bin} \approx 1$ and thus $P_0 \approx e^{-1} \approx 0.37$.
This probability does not depend on the bin width, so narrower bins do not correct this discrepancy.
Thus the expansion opacity can overestimate $P_0$ for a packet that traverses a wavelength bin with optically thick lines, compared to the line-by-line treatment.

Second, we compare the mean number of interactions of a packet in the bin.
The discrepancy in $P_0$ does not equate to a significantly smaller mean number of interactions with the expansion opacity than with the line-by-line treatment.
We show this with the example of a bin that contains one strong line with $\tau_l = 20$, and for simplicity we introduce the additional assumption of resonant scattering, i.e., that each line absorption event is followed by spontaneous re-emission in the same line.
An interaction then does not change the comoving wavelength of the packet, only its direction of propagation.
We count the number of interactions $N$ of one packet as it Doppler shifts across the bin, and we denote the mean number of interactions over many such packets by $\bar{N}$.
In the line-by-line treatment, each packet meets the line once, at its Sobolev resonance point.
There, the packet has one interaction with probability $1-e^{-\tau_l}$, and no interaction with probability $e^{-\tau_l}$.
After the interaction, the comoving wavelength of the packet increases beyond the line wavelength, so the packet cannot meet the line again.
Thus $N$ is either 0 or 1.
For $\tau_l = 20$, almost every packet has $N = 1$, so $\bar{N} \approx 1$.
Only a fraction $e^{-20} \approx 2.1\times10^{-9}$ of the packets crosses the bin with $N = 0$.

With a binned opacity, the opacity is constant across the bin, so the packet can interact at any point on its path in the bin.
After each interaction, the probability of a further interaction per unit path length stays the same.
Thus $N$ follows a Poisson distribution with mean $\bar{N} = \tau_\mathrm{bin}$.
The fraction of packets with $N = 0$ is $e^{-\tau_\mathrm{bin}}$, which is $P_0$ of equation~(\ref{eq:ptr}).
A Poisson distribution has only one parameter, so $\tau_\mathrm{bin}$ sets both $\bar{N}$ and $P_0$.
The expansion opacity gives $\tau_\mathrm{bin} = 1-e^{-\tau_l} \approx 1$ for our chosen $\tau_l = 20$.
Thus it gives $\bar{N}$ equal to the line-by-line value.
However, a Poisson distribution with mean 1 gives $N = 0$ to 37~per cent of the packets, $N = 1$ to 37~per cent, and $N \ge 2$ to 26~per cent.
Compared to the line-by-line treatment, the expansion opacity thus has the correct mean number of interactions, but it does not reproduce the distribution of the number of interactions.

In contrast, while the uncapped line-binned opacity gives the same value of $P_0$ as the line-by-line treatment, its mean number of interactions for our case of $\tau_l = 20$ with pure resonant scattering is $\bar{N} = \tau_\mathrm{bin} = 20$, not 1.
Thus a Monte Carlo simulation with the line-binned opacity could become significantly more expensive, since more interactions may need to be simulated when optically thick lines are present.
These additional interactions are not erroneous because they correspond to a physical process.
In a strong line, the Sobolev escape probability is approximately $1/\tau_l$, so, conceptually, a photon that is scattered by the line interacts approximately $\tau_l$ times near the resonance point before it escapes from the resonance region.
In our line-by-line treatment, these repeated interactions are taken into account analytically by treating the line interaction as one effective event, in which the transition rates are adjusted to account for multiple interactions with the Sobolev escape probability formalism \citep[see, e.g.,][]{lucy_monte_2002}.
Numerically, we thus need to simulate only one Monte Carlo event, but this event describes multiple scatterings at the resonance point.
The uncapped line-binned opacity instead follows a similar number of scatterings, but it spreads them across the bin.

In summary, and generalising to a bin with several lines that all scatter resonantly, the line-by-line treatment gives $\bar{N} = \sum_i (1-e^{-\tau_i})$ and $P_0 = \exp(-\sum_i \tau_i)$.
For weak lines, $1-e^{-\tau_i} \approx \tau_i$, so the line-by-line treatment gives $P_0 \approx e^{-\bar{N}}$, and both binned opacities match both values.
If the bin contains a strong line, a binned opacity can match only one of these two values with its single value for $\tau_\mathrm{bin}$.
The expansion opacity matches $\bar{N}$, and the uncapped line-binned opacity matches $P_0$ (Table~\ref{tab:binprob}).
With the expansion opacity, too many packets therefore cross a bin without an interaction wherever the bin contains a small number of optically thick lines.
In a bin with many optically thick lines, $\tau_\mathrm{bin}$ is large also for the expansion opacity, so almost every packet interacts.
With optically thick lines, the uncapped line-binned opacity still gives the same value for $P_0$ as the line-by-line case, but can lead to a larger number of interactions being simulated because it does not use a single effective scattering to replace multiple interactions at the resonance points.
The larger number of interactions per packet can give the uncapped line-binned opacity a high computational cost, as illustrated by the simple examples above.

We note that the illustrative $\bar{N}$ calculations presented above rely on the assumption of resonant scattering in all lines, which is not generally used in full simulations.
This assumption is nevertheless useful for illustrative purposes, because it separates the effect of the opacity treatment from that of the emission treatment, and $P_0$ does not depend on it.
The result for $\bar{N}$ also holds qualitatively in the general case, in which a packet is re-emitted at a different wavelength.
In our line-by-line treatment, the macroatom method \citep{lucy_monte_2002} still treats the repeated interactions at each resonance point analytically, for the line that absorbs the packet and for each transition that follows.
With a binned opacity, the packet is re-emitted at a random wavelength within its new bin (Section~\ref{sec:methodsemission}).
The packet then has to cross only a fraction $\xi$ of the optical depth of this bin, where $\xi$ is uniformly distributed between 0 and 1, and it does so without a further interaction with probability $e^{-\xi\tau_\mathrm{bin}}$.
Averaged over $\xi$, this probability is $(1-e^{-\tau_\mathrm{bin}})/\tau_\mathrm{bin} \approx 1/\tau_\mathrm{bin}$ for $\tau_\mathrm{bin} \gg 1$, so the number of interactions again increases with $\tau_\mathrm{bin}$.
In the line-by-line treatment with thermal redistribution (the LBL + thermal configuration of Section~\ref{sec:methodsemission}), the re-emitted packet also starts at a random wavelength in its bin.
In a bin with one strong line, this packet leaves the bin without a further interaction if it starts redward of the line, and for a random line position within the bin the probability of this is one half, independent of $\tau_l$.

\begin{table}
\caption{Transport of a packet across one wavelength bin that contains lines $i$ with Sobolev optical depths $\tau_i$, for each opacity treatment and for a bin of any width.
The second column gives the optical depth across the bin, $\tau_\mathrm{bin}$ (equation~\ref{eq:taubin}).
The last two columns are for a bin with one line of optical depth $\tau_l = 20$.
The third column gives the mean number of interactions $\bar{N}$ of a packet that enters the bin, if an interaction does not change the comoving wavelength of the packet (resonant scattering).
The fourth column gives $P_0 = e^{-\tau_\mathrm{bin}}$, which is the probability that a packet crosses the bin without an interaction.}
\label{tab:binprob}
\centering
\begin{tabular}{lccc}
\hline\hline
Treatment & $\tau_\mathrm{bin}$ & $\bar{N}$ & $P_0$\\
 & & \multicolumn{2}{c}{(one line, $\tau_l = 20$)}\\
\hline
Line-by-line                           & $\sum_i \tau_i$ & 1.00 & $2.1\times10^{-9}$\\
Line-binned, uncapped                  & $\sum_i \tau_i$ & 20.0 & $2.1\times10^{-9}$\\
Line-binned, $\tau_\mathrm{l,max}=10$  & $\sum_i \min(\tau_i, 10)$ & 10.0 & $4.5\times10^{-5}$\\
Line-binned, $\tau_\mathrm{l,max}=5$   & $\sum_i \min(\tau_i, 5)$ & 5.00 & 0.007\\
Line-binned, $\tau_\mathrm{l,max}=1$   & $\sum_i \min(\tau_i, 1)$ & 1.00 & 0.37\\
Expansion opacity                      & $\sum_i \left(1-e^{-\tau_i}\right)$ & 1.00 & 0.37\\
\hline
\end{tabular}
\tablefoot{For the binned opacities, the optical depth of each line is spread evenly across the bin, so $\bar{N} = \tau_\mathrm{bin}$ and $P_0 = e^{-\bar{N}}$.
In the line-by-line treatment, the packet meets each line at its resonance point, and $\tau_\mathrm{bin}$ is the sum of the optical depths of the lines.
In this treatment, $\bar{N} = \sum_i (1-e^{-\tau_i})$.
The cap $\tau_\mathrm{l,max}$ applies to each line (equation~\ref{eq:binnedopac}), not to the bin.
The uncapped line-binned opacity is that of \citet{fontes_linebinned_2020}.
Real bins contain $10^3$--$10^5$ transitions (Fig.~\ref{fig:expansionopacitycomparison}), most of which are optically thin, and the case of one line shows the effect of a single optically thick line.}
\end{table}

\subsubsection{Capped line-binned opacities}
\label{sec:cappedopac}
In ARTIS, the explicit re-absorptions make the Monte Carlo transport with the uncapped line-binned opacity computationally challenging.
In particular, the assumed Kirchhoff emissivity sampling (equation~\ref{eq:kirchhoff}) favours injecting packets into high-opacity bins in which the probability of significant propagation before re-absorption is small.
A packet then cycles between absorption and re-emission many times while it barely moves in space or time, so the calculation uses computing time while hardly advancing the simulation time.
The number of these cycles is large when many bins are optically thick.
A code with a diffusion treatment for optically thick cells, such as the hybrid implicit and discrete diffusion Monte Carlo scheme of SuperNu \citep{wollaeger_radiation_2013,wollaeger_broad_2021}, handles such conditions efficiently, because it does not follow every interaction in the optically thick cells, whereas ARTIS does.
We therefore put a cap $\tau_\mathrm{l,max}$ on the contribution of each line to the sum of equation~(\ref{eq:linebinnedopac}):
\begin{align}
    \kappa_{\mathrm{binned}} &= \frac{1}{ct\rho} \sum_i \frac{\lambda_i}{\Delta\lambda} \min\left(\tau_i, \tau_\mathrm{l,max}\right).
    \label{eq:binnedopac}
\end{align}
A test calculation with the uncapped line-binned opacity did not complete an early timestep within 24 hours of wall time on 2000 cores, compared with 1 hour per timestep with $\tau_\mathrm{l,max} = 1$.

For a bin with one strong line of optical depth $\tau_l > \tau_\mathrm{l,max}$, the capped line-binned opacity gives $\tau_\mathrm{bin} = \tau_\mathrm{l,max}$.
Such a bin then has $P_0 = e^{-\tau_\mathrm{l,max}}$ (equation~\ref{eq:ptr}) and, for resonant scattering, a mean number of interactions of $\bar{N} = \tau_\mathrm{l,max}$.
A higher cap thus brings $P_0$ closer to the line-by-line value (Table~\ref{tab:binprob}), but it raises the cost.
We increase the cap from $\tau_\mathrm{l,max} = 1$ to 5 and 10 to test the convergence towards the uncapped line-binned opacity.
If the results with $\tau_\mathrm{l,max} = 5$ and 10 are similar, the uncapped line-binned opacity of \citet{fontes_linebinned_2020} is likely to give similar results (Section~\ref{sec:resultsopac}).

The lowest cap, $\tau_\mathrm{l,max} = 1$, gives an opacity close to the expansion opacity.
This is clear if we compare the contributions of one line to the two opacities, $\min(\tau_i, 1)$ and $1-e^{-\tau_i}$.
For $\tau_i \ll 1$, both terms are approximately $\tau_i$.
For $\tau_i \gg 1$, both terms are approximately one.
Their ratio, $\min(\tau_i, 1)/(1-e^{-\tau_i})$, increases with $\tau_i$ for $\tau_i < 1$ and decreases for $\tau_i > 1$.
The ratio has its maximum at $\tau_i = 1$, where it is $e/(e-1) \approx 1.58$.
The contribution of each line to this capped line-binned opacity is therefore between 1 and 1.58 times its contribution to the expansion opacity.
For the same level populations, the line-binned opacity with $\tau_\mathrm{l,max} = 1$ thus lies between $\kappa_{\mathrm{exp}}$ and $1.58\,\kappa_{\mathrm{exp}}$ in each bin.
Like the expansion opacity, this capped opacity saturates for strong lines (Table~\ref{tab:binprob}).

\subsubsection{Implementation and comparison of the binned opacities}
\label{sec:binnedimpl}
In this work, we use bins with constant $\Delta\lambda = 20$~\AA\ from 60 to 40\,000~\AA\ (1997 bins).
This range extends to much shorter wavelengths than the 1024 logarithmically spaced wavelength groups from 1000 to 128\,000~\AA\ in the simulations of \citet{fontes_linebinned_2020}.
We include the short wavelengths to capture the trapping of radiation during the first day, when the hot, highly ionised ejecta interact mainly with ultraviolet radiation \citep[cf.][]{banerjee_simulations_2020}.
Only the bound--bound opacity is binned.
Electron scattering and free--free opacities are evaluated at the packet frequency as in the line-by-line calculation.
The binned opacities are recomputed for each cell at each timestep from the LTE level populations.
ARTIS uses the relativistic Doppler shift, so the true path length across a bin differs from $\Delta s$ by a factor that depends on the position and direction of the packet.
In the four binned-opacity configurations (Section~\ref{sec:methodsemission}), ARTIS therefore multiplies the binned opacity by the ratio of $\Delta s$ to the true path length.
With this correction, the optical depth across each bin is equal to the value of equation~(\ref{eq:taubin}).

\begin{figure}
\centering
\includegraphics[width=\hsize]{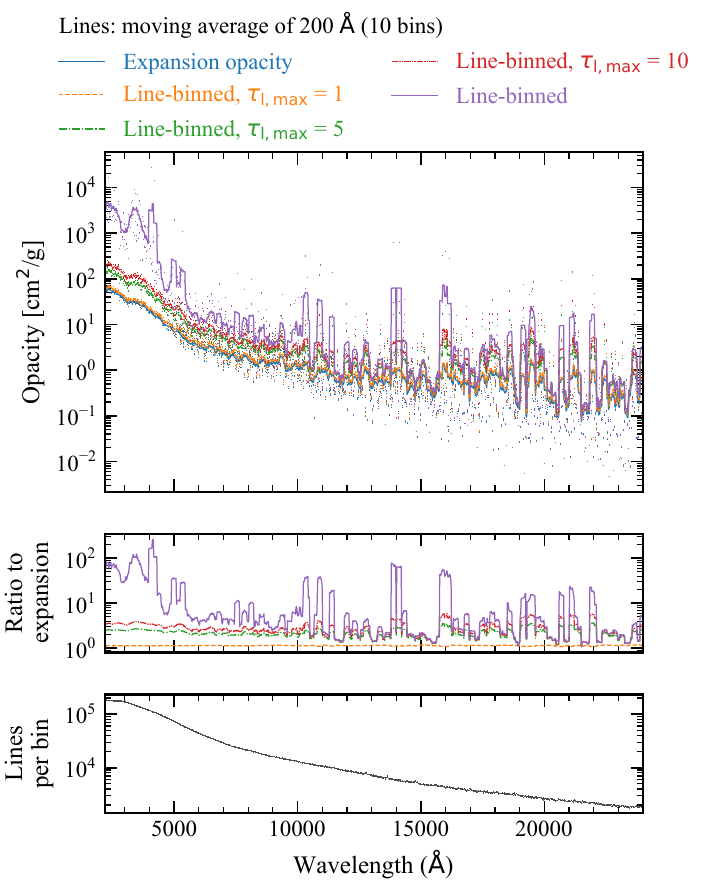}
   \caption{Binned line opacities of the 3D dyn 8x mass v/2 model at 1.4 days, averaged over all cells with the cell masses as weights: the expansion opacity, the capped line-binned opacities with $\min(\tau_i, \tau_\mathrm{l,max})$ for $\tau_\mathrm{l,max} = 1$, 5, and 10, and the uncapped line-binned opacity with $\tau_i$ in place of $\min(\tau_i, \tau_\mathrm{l,max})$ (labelled `Line-binned').
The level populations are LTE populations from the temperatures and ion populations of the line-by-line (LBL) macroatom calculation (Section~\ref{sec:methodsemission}).
In the top panel, the short horizontal lines give the individual 20~\AA\ bins, and the thick lines give a centred moving average over 10 bins (200~\AA).
The middle panel gives the ratio of each moving average to that of the expansion opacity.
The bottom panel gives the number of transitions that ARTIS includes in each bin.}
     \label{fig:expansionopacitycomparison}
\end{figure}

Figure~\ref{fig:expansionopacitycomparison} compares the binned opacities of the velocity-compressed model at 1.4 days, each averaged over all cells with the cell masses as weights.
The line-binned opacity with $\tau_\mathrm{l,max} = 1$ is $\sim$1.1--1.2 times the expansion opacity at all wavelengths, within the bounds of 1 and 1.58.
With $\tau_\mathrm{l,max} = 5$ and 10, the opacity is $\sim$1.5--5 times the expansion opacity.
With the uncapped line-binned opacity, the opacity is $\sim$30--250 times the expansion opacity below $\sim$4500~\AA, where many lines are optically thick.

\subsection{Emission treatment and model configurations}
\label{sec:methodsemission}
Once the lines are binned, the transition that absorbed a given packet is no longer known, so the emission that follows each excitation cannot be calculated in detail.
The fluorescence cascades of individual transitions, in which an absorption in one line leads to emission in other lines, are therefore lost, and codes that use binned opacities typically assume that the absorbed radiation is fully or mostly redistributed thermally at each interaction \citep[e.g.,][]{wollaeger_impact_2018,bulla_possis_2019}.
Thermal redistribution means that the re-emitted radiation has a spectrum set by the local temperature and opacity of the gas, with no memory of the transition that absorbed it.
The emissivity $j_\lambda$ is then commonly taken to be the product of the Planck function $B(\lambda, T_e)$, the mass density, and the true absorption opacity, which excludes scattering (Kirchhoff's law).
Such an emissivity is assumed, for example, by \citet[][equation 11]{kasen_timedependent_2006} and \citet[][equation 6]{tanaka_radiative_2013}.

In our standard treatment, the emission after each absorption is instead sampled with the macroatom method of \citet{lucy_monte_2002,lucy_monte_2003}.
A macroatom has the energy levels of a real ion, but it absorbs and emits whole packets.
When a packet is absorbed in a line, the packet becomes a macroatom in the upper level of that line.
The macroatom then makes a sequence of transitions between its levels, upward or downward, by radiative and collisional processes.
ARTIS selects each transition at random, with a probability proportional to the rate at which the transition moves energy.
The sequence ends when the macroatom emits the packet in a line, which is in general not the line that absorbed the packet (fluorescence), or when a collision converts the energy of the packet into thermal energy of the gas.
A packet of thermal energy, from this conversion or from the deposited decay energy, becomes radiation again by free--free emission or becomes a new macroatom by a collisional excitation.
For given transition rates, the macroatom method needs no assumption about the emission spectrum, because it samples each transition with the probability that follows from these rates \citep{lucy_monte_2002}.
Thermal redistribution instead takes the emissivity of LTE (equation~\ref{eq:kirchhoff}), which is exact only if collisions set the level populations or if the radiation field is close to a Planck function at the gas temperature.
Neither of these two conditions is guaranteed in the ejecta, and the macroatom method does not need them for the emission, because it uses the LTE populations only for the opacities and the transition rates.
We therefore regard the line-by-line macroatom calculation as our most accurate one (Section~\ref{sec:resultsopac}).
The disadvantages of the macroatom method are its cost and its need for the rates of all transitions from the current level at each step, which a binned treatment avoids.

To quantify the effect of the binned opacities and of thermal redistribution, we compare six configurations of the same 3D model (Section~\ref{sec:resultsopac}).
The baseline is our line-by-line (LBL) macroatom calculation (`LBL + macroatom'), while the five alternatives all redistribute the absorbed radiation thermally at each line interaction and differ only in their opacities.
These opacities are the line-by-line opacity (`LBL + thermal'), the expansion opacity (`Expansion + thermal'), and the capped line-binned opacities with $\tau_\mathrm{l,max} = 1$, 5, and 10 (`Binned ($\tau_\mathrm{l,max}=1$) + thermal', `Binned ($\tau_\mathrm{l,max}=5$) + thermal', and `Binned ($\tau_\mathrm{l,max}=10$) + thermal').
We use `thermal redistribution' for this treatment of the absorbed radiation, and we reserve `thermalisation' for the deposition of the energy of the radioactive decay products in the gas (Section~\ref{sec:rt}).

In all five thermal-redistribution configurations, ARTIS samples the wavelength bin of each re-emitted packet from the Kirchhoff emissivity
\begin{align}
   j_\lambda &= \rho\,B(\lambda,T_e)\,\left[\kappa_\mathrm{bb}(\lambda) + \kappa_\mathrm{ff}(\lambda)\right],
   \label{eq:kirchhoff}
\end{align}
where $\kappa_\mathrm{ff}$ is the free--free opacity and $\kappa_\mathrm{bb}$ is the binned line opacity, both evaluated at the centre of each bin.
The binned line opacity is $\kappa_\mathrm{binned}$ (equation~\ref{eq:binnedopac}) with the corresponding value of $\tau_\mathrm{l,max}$ in the three Binned + thermal configurations, and $\kappa_\mathrm{exp}$ (equation~\ref{eq:expopac}) in the LBL + thermal and Expansion + thermal configurations.
The LBL + thermal configuration uses $\kappa_\mathrm{exp}$ only in this emissivity, and its transport stays line-by-line.
ARTIS re-emits the thermal energy from all other sources, such as the deposited decay energy, with the same distribution.
Within the selected bin, ARTIS samples the frequency of the packet uniformly, so the wavelength of the packet is random within the bin.
In the LBL + thermal configuration, a re-emitted packet that starts blueward of an optically thick line in its bin can therefore be absorbed again by that line.
A line-by-line macroatom emissivity puts the emission at the line wavelengths and does not have this additional self-absorption.

\subsection{Comparison data for AT2017gfo}
\label{sec:obsdata}
We compare our models with the Very Large Telescope X-shooter spectra of AT2017gfo published by \citet{pian_spectroscopic_2017} and \citet{smartt_kilonova_2017} in the flux-calibrated reduction of the ENGRAVE data release.\footnote{\url{http://www.engrave-eso.org/AT2017gfo-Data-Release/}}
The spectra of this data release are corrected for Galactic extinction and shifted to the rest frame of the host galaxy NGC~4993 (redshift $z \approx 0.01$).
All wavelengths in this work are rest-frame wavelengths, and the model fluxes at 40~Mpc need no further correction for this small redshift.
We adopt a distance of 40~Mpc, which is consistent with the gravitational-wave distance \citep{abbott_gw170817_2017} and with the published spectral reductions.
The distance of 40.7~Mpc from surface brightness fluctuations \citep{cantiello_precise_2018} has an uncertainty of $\sim$6~per cent, which corresponds to $\sim$12~per cent in the model fluxes.

We quote epochs as rest-frame times since the merger, which is also the reference time of our model output.
We focus on the epochs at 1.4, 2.4, and 3.4 days, which sample the early sequence of spectra.
Spectra before 1.4 days exist \citep[see the compilation of][]{sneppen_emergence_2024}, but they are optical spectra from other instruments without near-infrared coverage.
We therefore restrict the spectral comparison to the X-shooter sequence, which covers the ultraviolet to the near-infrared with one instrument.
The dependence of the early luminosity on the snapshot thermal energy (Section~\ref{sec:rt}) is a second reason to begin the spectral comparison at 1.4 days.

For the bolometric comparison, we use the ultraviolet--optical--infrared (UVOIR) luminosity estimates for AT2017gfo of \citet{smartt_kilonova_2017} and \citet{waxman_constraints_2018}.
\citet{smartt_kilonova_2017} fitted a blackbody to the photometry and integrated the fit from 1000 to 25\,000~\AA, whereas \citet{waxman_constraints_2018} integrated the multiband photometry and added a Rayleigh--Jeans extrapolation redward of the reddest band.
The two bolometric estimates differ by up to 40~per cent before 1.5 days.

\section{Results for scaled models}
\label{sec:resultsscaled}
\subsection{Bolometric luminosity evolution}
\label{sec:lightcurves}

\begin{figure*}
     \begin{center}
     \includegraphics[width=0.95\hsize]{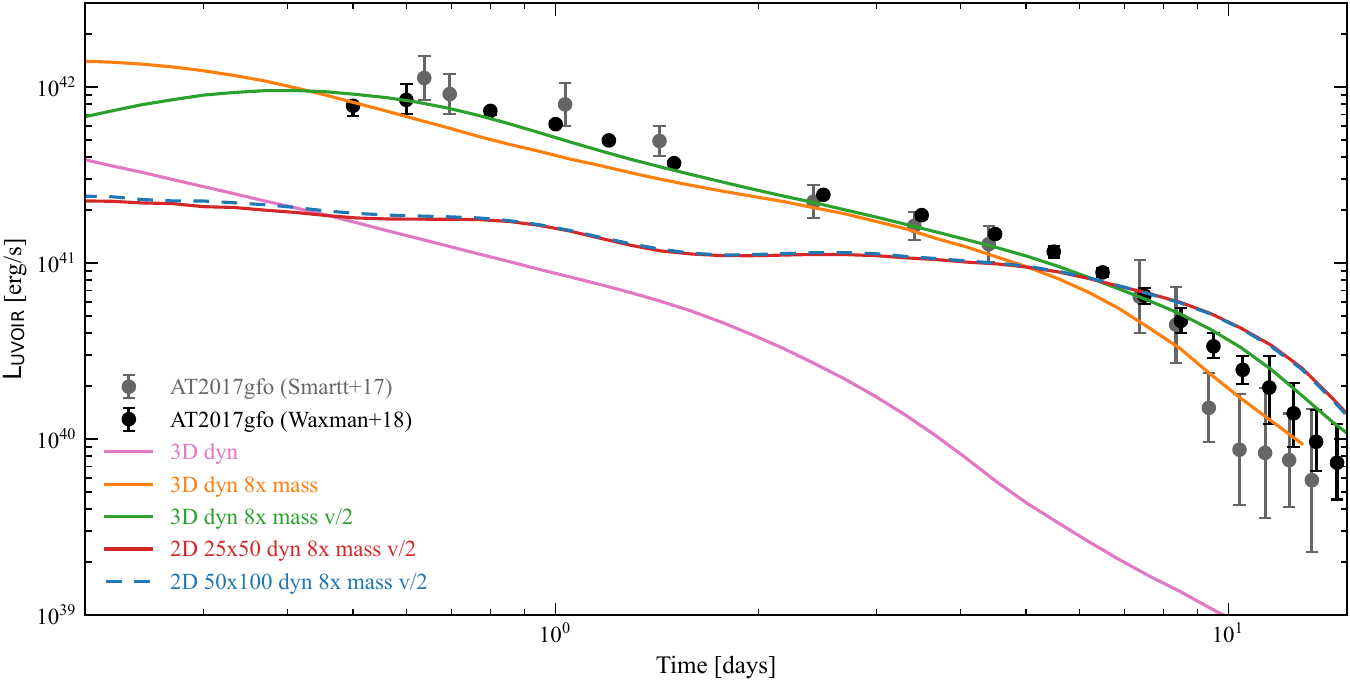}
     \end{center}
   \caption{Isotropic-equivalent bolometric luminosity (ultraviolet--optical--infrared power) in the polar direction for the unscaled 3D model, the two scaled 3D models, and the 2D models on the $25{\times}50$ and $50{\times}100$ grids, compared with bolometric estimates for AT2017gfo \citep{smartt_kilonova_2017,waxman_constraints_2018}.}
     \label{fig:lightcurves}
\end{figure*}

Figure~\ref{fig:lightcurves} shows the bolometric light curves of our models together with the luminosity estimates for AT2017gfo described in Section~\ref{sec:obsdata}.
As in \citet{shingles_selfconsistent_2023}, the unscaled model is almost an order of magnitude fainter than AT2017gfo at 1 day and declines far more steeply than the observations thereafter.
Mass scaling raises the luminosity to within a factor of two of the observations between 0.5 and 1.5 days, and to within $\sim$30~per cent between 1.5 and 5 days.
Between $\sim$5 and 8 days, the mass-scaled model remains $\sim$30~per cent fainter than the observations.

Halving the velocities leaves the mass of radioactive material unchanged but raises the density, so a larger fraction of the decay energy is deposited in the ejecta.
We find that the thermalisation efficiency of the $\gamma$-rays at 1.4 days rises from $\sim$0.3 in the mass-scaled model to $\sim$0.55 in the velocity-compressed model, and that the total deposited power is $\sim$30~per cent higher between 1 and 4 days.
The $\beta$ particles are almost fully thermalised in both models at these epochs.
The higher density also traps the UVOIR radiation for longer, which delays the emission.
As a consequence, the velocity-compressed model is fainter than the mass-scaled model before $\sim$0.4 days but brighter at all later times.

Between 0.8 and 1.5 days, the velocity-compressed model is $\sim$10--20~per cent fainter than the lower of the two observational estimates.
The snapshot thermal energy supplies about a quarter of the model luminosity at these epochs.
Since this energy is an overestimate (Section~\ref{sec:rt}), a correction would make the model fainter.
The velocity-compressed model tracks the observed decline between $\sim$2 and 8 days and is brighter than the observations after $\sim$8 days.
At these late epochs, the model still releases radiation that it trapped earlier, and we find that its luminosity, averaged over all directions, exceeds the deposited power by 50--80~per cent between 6 and 12 days.

\subsection{Spectral evolution}
\begin{figure*}
     \begin{center}
     \includegraphics[width=0.9\hsize]{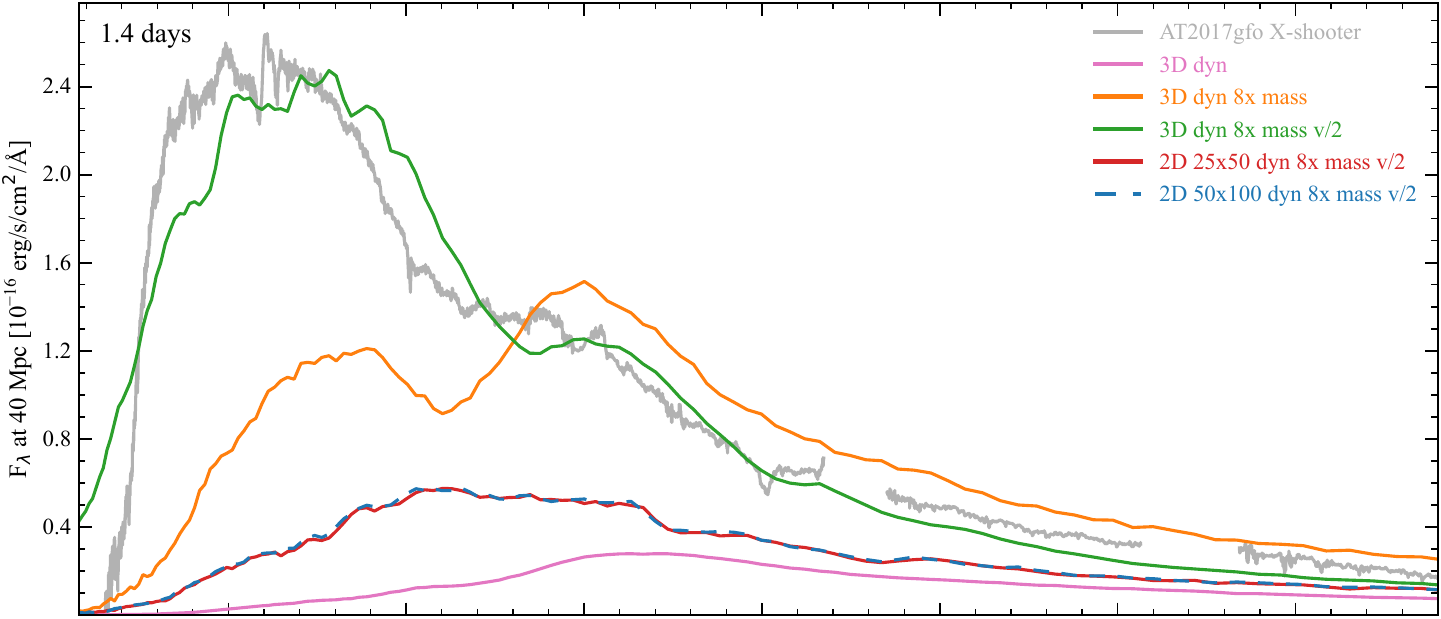}
     \includegraphics[width=0.9\hsize]{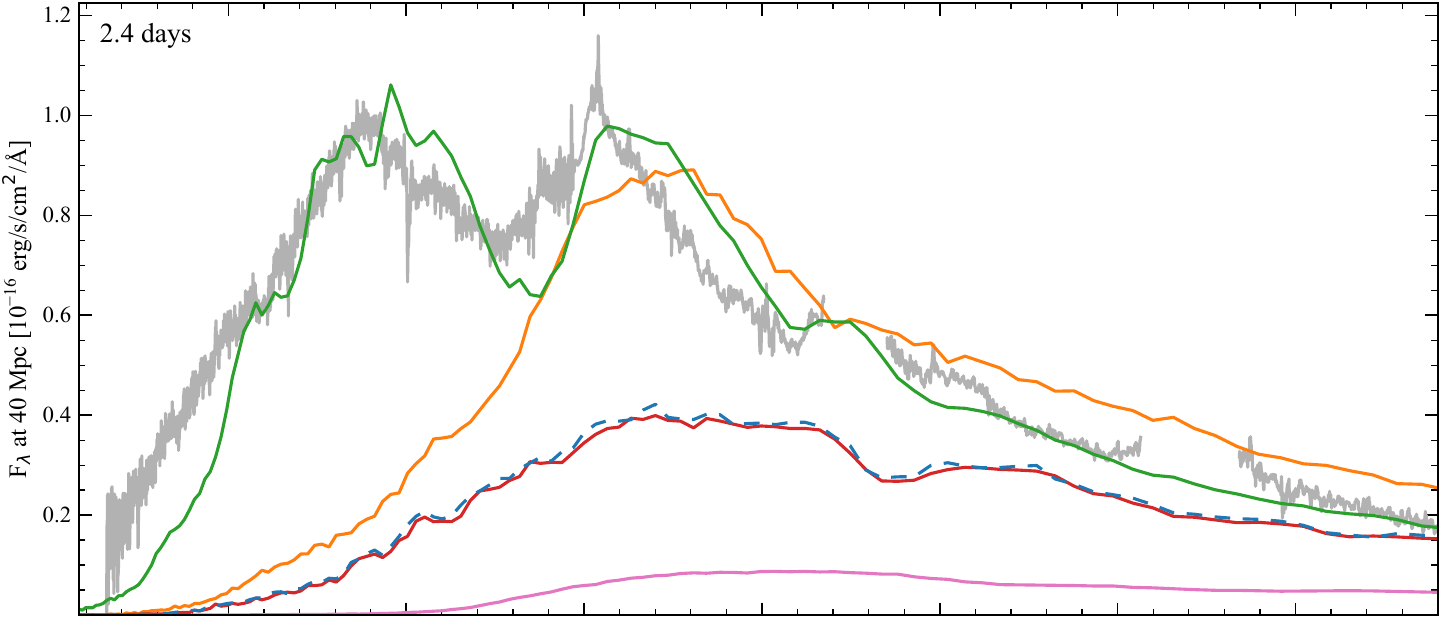}
     \includegraphics[width=0.9\hsize]{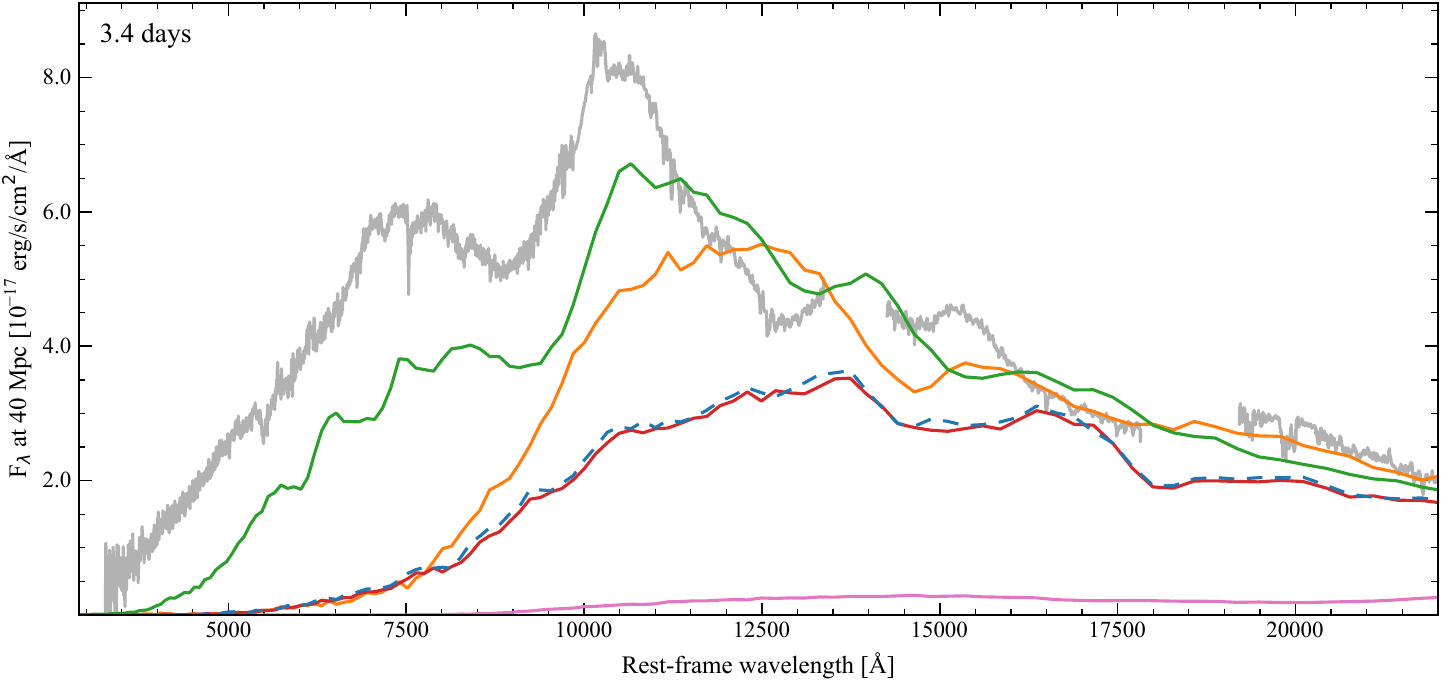}
     \end{center}
     \caption{Spectra in the polar direction at 1.4 days (top), 2.4 days (middle), and 3.4 days (bottom), for the unscaled 3D, scaled 3D, and 2D models, compared with the observed AT2017gfo spectra \citep{pian_spectroscopic_2017,smartt_kilonova_2017} at a distance of 40~Mpc.
The wavelengths are in the rest frame of the host galaxy.}
     \label{fig:spectra_series}
\end{figure*}

Figure~\ref{fig:spectra_series} compares the model spectra at 1.4, 2.4, and 3.4 days with the X-shooter spectra of AT2017gfo (Section~\ref{sec:obsdata}).
Increasing the ejecta mass alone gives spectra that are too red at early times.
At 1.4 days, the mass-scaled model peaks near 10\,000~\AA\ with a strong flux deficit at blue wavelengths, whereas the observed spectrum peaks near 5500~\AA.
Halving the velocities as well moves the spectral peak blueward and raises the optical flux, especially at 1.4 and 2.4 days.
At 1.4 days, the velocity-compressed 3D model matches the observed peak flux and reproduces the shape of the red and near-infrared tail.
However, its peak lies near 6500~\AA, about 1000~\AA\ redward of the observed peak, and the model underestimates the flux blueward of $\sim$5500~\AA.
Between $\sim$11\,000 and 15\,000~\AA\ at this epoch, the velocity-compressed model follows the observed flux closely, while it is fainter than observed redward of $\sim$15\,000~\AA.

At 2.4 days, the velocity-compressed 3D model reproduces the observed double-peaked structure and flux level, with a broad optical peak near 7000~\AA\ and a near-infrared peak near 10\,500~\AA, both close to the observed wavelengths.
However, the trough between the peaks (9000--10\,000~\AA) is deeper than observed, and the flux blueward of $\sim$5500~\AA\ is underestimated.
At 3.4 days, the velocity-compressed 3D model remains the closest match, peaking close to the observed peak near 10\,500~\AA\ but falling short of the observed peak flux, and all models underestimate the observed flux in the optical.

At 1.4 and 2.4 days, the velocity-compressed model reproduces the observed flux level between 5500 and 15\,000~\AA, apart from the trough near 9500~\AA\ and an excess on the red side of the near-infrared peak at 2.4 days.
The optical deficit redward of 5500~\AA\ thus develops after 2.4 days, and departures from our assumption of LTE could contribute to this deficit at 3.4 days (Section~\ref{sec:limitations}).
At 3.4 days, the model is redder than observed rather than uniformly fainter, as it gives about half of the observed flux between 5500 and 7500~\AA\ but about 80~per cent between 5500 and 15\,000~\AA.

\subsection{Spectral feature contributions}
\label{sec:featureid}
We associate spectral features with individual ions and transitions by recording, for each escaping Monte Carlo packet, its last interaction before it reached the observer.
The last emission event sets the wavelength at which the packet escapes, and we also record the last absorption before that emission and its wavelength.
The emission flux of an ion or a transition is the flux of the escaping packets whose last emission was in that ion or transition.
The absorption flux is the flux of the escaping packets whose last absorption was in that ion or transition.
Both fluxes are last-interaction quantities of the escaping packets only, and they do not measure the total emission and absorption in the ejecta, most of which happens in packets that interact again before they escape.
This method does not test how the spectrum changes when the opacity of one ion is removed from the calculation, so by itself it cannot establish a unique identification in the observed spectrum.
Such an identification requires fits to the observed spectra in which the composition is varied and species are added or removed, as \citet{gillanders_modelling_2022,gillanders_modelling_2024} did for the full sequence of spectra of AT2017gfo.

\subsubsection{Ion contributions}
\label{sec:ioncontributions}
\begin{figure*}
     \begin{center}
     \includegraphics[width=\hsize]{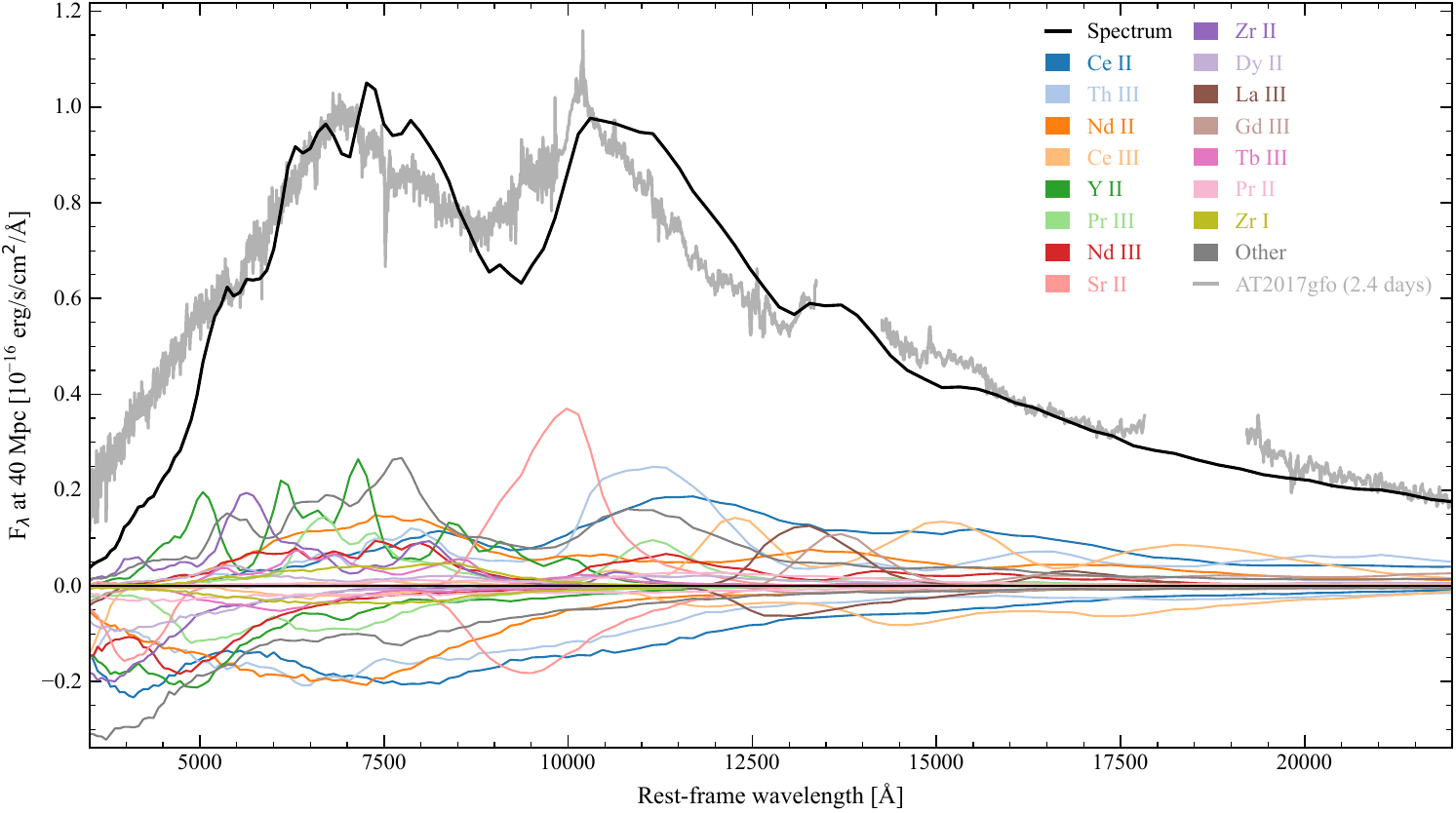}
     \includegraphics[width=\hsize]{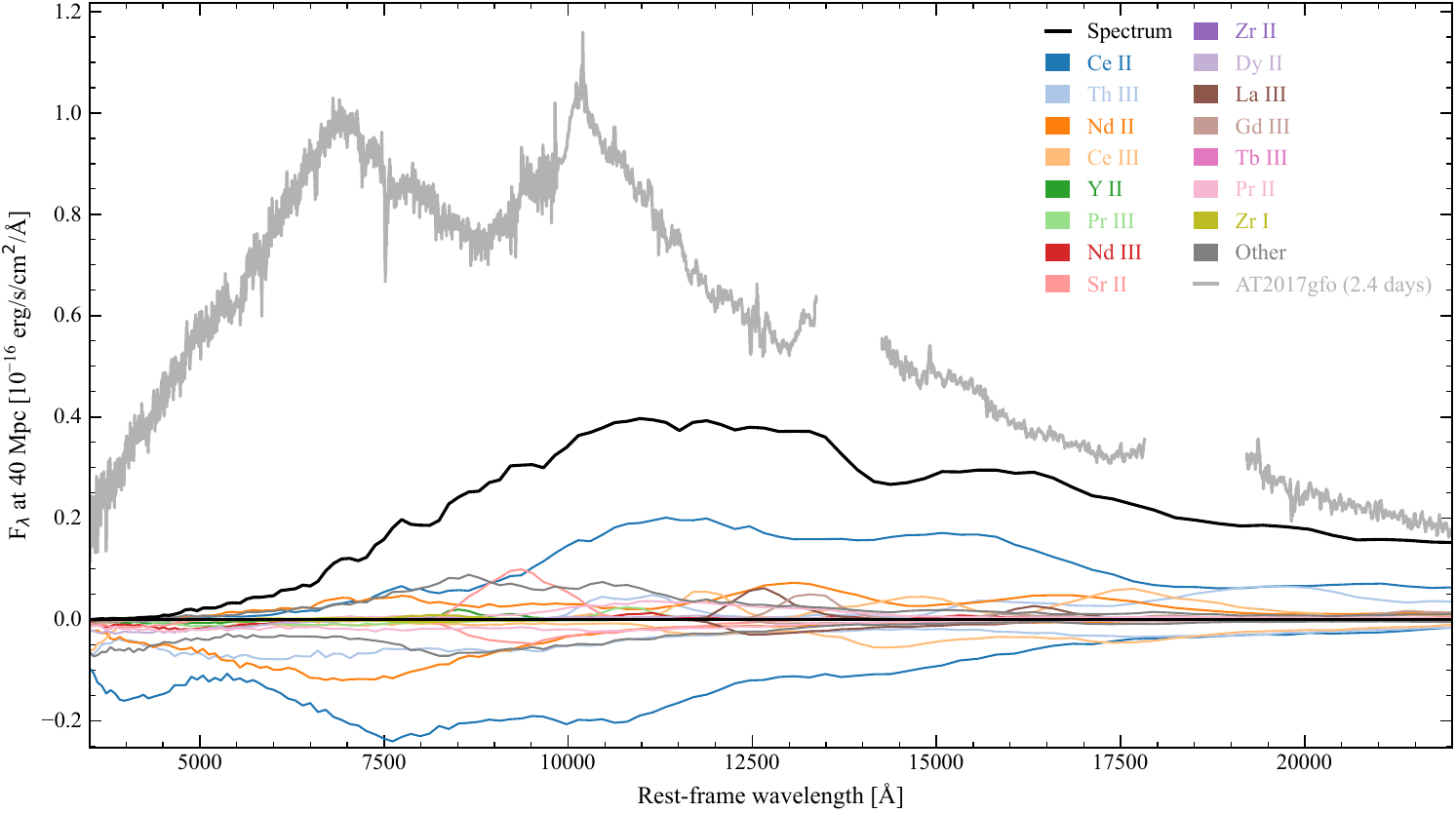}
     \end{center}
   \caption{Spectra in the polar direction at 2.4 days for the dyn 8x mass v/2 model in 3D (top) and its 2D counterpart on the $25{\times}50$ grid (bottom), decomposed by the ion responsible for the last emission or absorption event of each escaping packet.
Positive contributions show the emission flux of each ion and negative contributions its absorption flux, both defined by the last interaction of the escaping packets (Section~\ref{sec:featureid}), not by the total emission and absorption in the ejecta.
The `Other' curve groups the contributions of all remaining ions, and the observed AT2017gfo spectrum at 2.4 days is shown in grey for comparison.}
     \label{fig:spectra_ions_2p4d_3d_dyn_mass08x_veldiv2}
\end{figure*}

Figure~\ref{fig:spectra_ions_2p4d_3d_dyn_mass08x_veldiv2} (top panel) shows this decomposition in the polar direction for the velocity-compressed 3D model at 2.4 days, an epoch that we choose because the observed double-peaked structure is well developed.
The absorption flux of \ion{Ce}{ii} and several other ions exceeds their emission flux at short wavelengths (below $\sim$10\,000~\AA\ for \ion{Ce}{ii}), and the reverse holds at longer wavelengths.
In the model, this pattern shows that these ions absorb radiation at short wavelengths and re-emit the energy at longer wavelengths, as expected for fluorescence.

\ion{Pr}{iii} produces an emission peak near 6700~\AA\ and a weaker one near 7300~\AA, within the broad optical peak of the spectrum.
Its emission flux is about three quarters of that of \ion{Sr}{ii}, and its absorption flux, which lies at shorter wavelengths than the emission, is almost as large as its emission flux.
\ion{Pr}{iii} therefore mainly redistributes escaping flux from the blue into the 6500--7500~\AA\ region by fluorescence.

The \ion{Pr}{iii} data in AD3 are the calibrated data of \citet{flors_calibrated_2026}.
The transitions that form the two peaks connect the $4f^3$ ground configuration with the low-lying $4f^2 5d$ levels, most of which have experimental energies, so the calibration fixes the wavelengths of the transitions that carry about 90~per cent of this emission.
In the uncalibrated atomic-structure data, most of the $4f^2 5d$ levels lie too high, and most of the same lines are shorter in wavelength by several hundred \AA.
The calibration therefore sets the position of the \ion{Pr}{iii} emission.
As the strength of this contribution depends on the completeness and the transition probabilities of the \ion{Pr}{iii} line list (Section~\ref{sec:transitions}), dedicated atomic-structure calculations for \ion{Pr}{iii} are needed to confirm the strength.
\citet{ford_kilonovae_2024} find \ion{Pr}{iii} among the strongest absorbing ions in synthetic spectra of low-$Y_e$ ejecta, and \citet{collins_exploring_2026} find \ion{Pr}{iii} contributions in the optical in some of their merger models but not in others.

\ion{Pr}{iii} forms the model peak near 6700~\AA, and \ion{Y}{ii} is the largest contributor to the stronger model peak near 7300~\AA, both within the broad optical peak of the model.
The observed spectrum at 2.4 days has one broad maximum near 6900~\AA, where the model has a local minimum between these two peaks (Fig.~\ref{fig:spectra_lines_2p4d_3d_dyn_mass08x_veldiv2}), so we do not identify either model peak with an observed feature.

\ion{Sr}{ii} makes the largest single-ion contribution to the persistent near-infrared peak near 10\,500~\AA, as in the merger models of \citet{collins_exploring_2026}.
\ion{He}{i} $\lambda$10830 has been proposed as an alternative or additional contributor to this feature \citep{perego_production_2022,tarumi_nonlte_2023}.
However, \citet{perego_production_2022} find that a significant \ion{He}{i} feature requires either a luminosity below that of AT2017gfo or a He mass at least ten times larger than their merger simulations predict.
\citet{sneppen_helium_2024} also find that the emergence time and the evolution of the feature are inconsistent with \ion{He}{i}.
Our LTE calculations cannot produce a \ion{He}{i} $\lambda$10830 contribution, because the metastable lower level of this line is populated by non-thermal excitation and recombination \citep{tarumi_nonlte_2023}, not by thermal excitation.

The \ion{Sr}{ii} peak is blended on its red side with \ion{Th}{iii} and \ion{Ce}{ii} emission, which peaks near 11\,000--11\,500~\AA, and with a combined contribution of similar size from the ions that the figure does not show individually (the `Other' curve).
\ion{Th}{iii} has the second-largest summed emission and absorption flux at this epoch, behind only \ion{Ce}{ii}.
\ion{Th}{iii} absorbs between $\sim$4000 and 10\,000~\AA\ and re-emits mainly between 10\,500 and 12\,500~\AA, which broadens the red wing of the near-infrared peak.
However, we treat this contribution with caution (Section~\ref{sec:limitations}) because the \ion{Th}{iii} atomic data are theoretical and not calibrated to measured wavelengths.
Th is also the only actinide with atomic data in AD3, and \ion{Th}{iii} is its only ionisation stage, which means that the opacity of the other actinides and of the other Th ions is missing.

Apart from its red wing, the \ion{Sr}{ii} contribution is narrower and less blended than the features at shorter wavelengths.
Because this feature comes mainly from one triplet with known rest wavelengths, its position depends primarily on the expansion velocity of its line-forming region, although blends broaden its red side.
At 2.4 and 3.4 days, the feature in the velocity-compressed model peaks slightly redward of the observed feature and is broader, mainly on its red side.
The mass-scaled model with the original velocities shows only a broad hump at this wavelength (Fig.~\ref{fig:spectra_series}).

\subsubsection{Individual transitions}
\label{sec:transitions}
\begin{figure*}
     \begin{center}
     \includegraphics[width=\hsize]{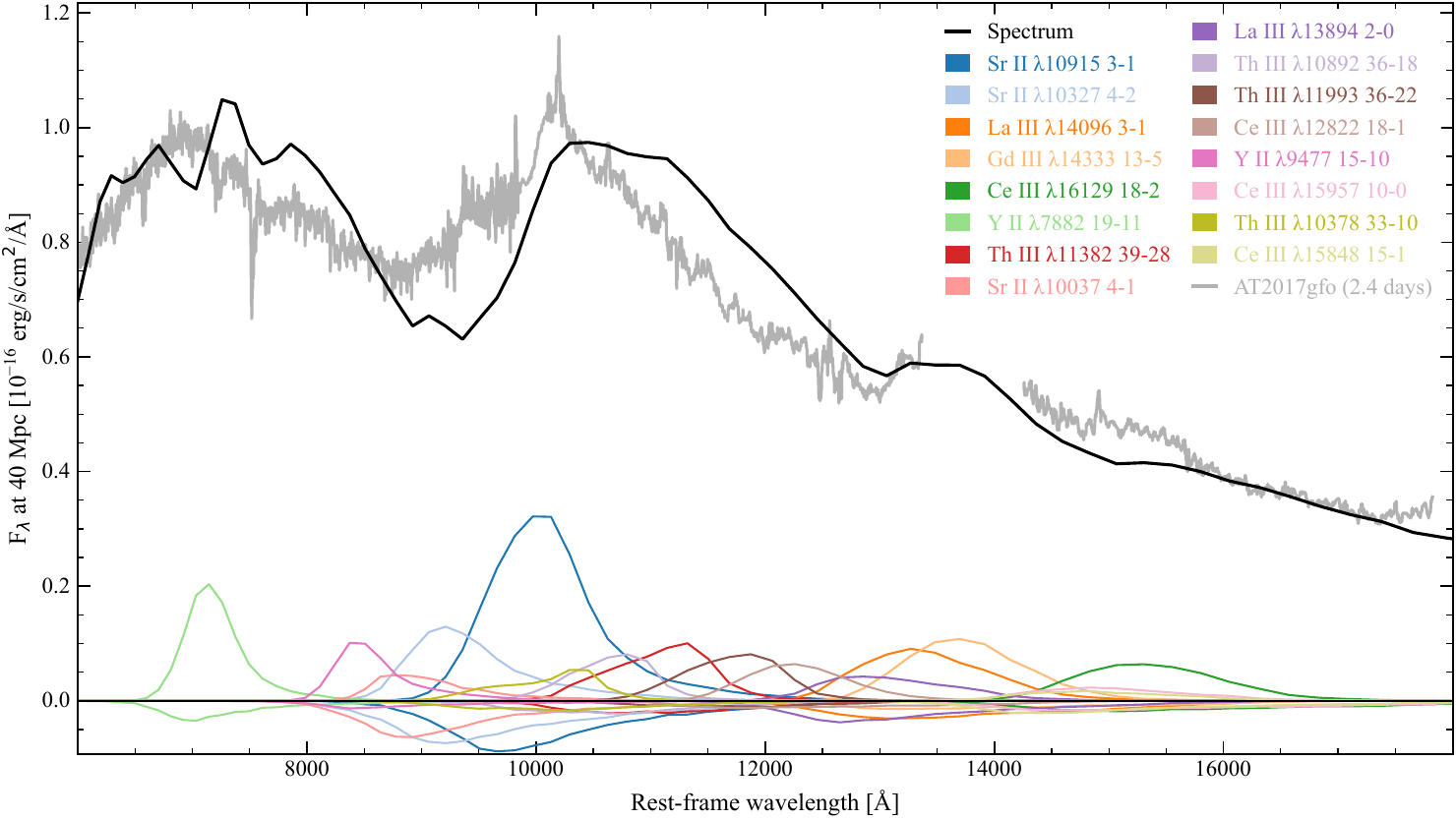}
     \includegraphics[width=\hsize]{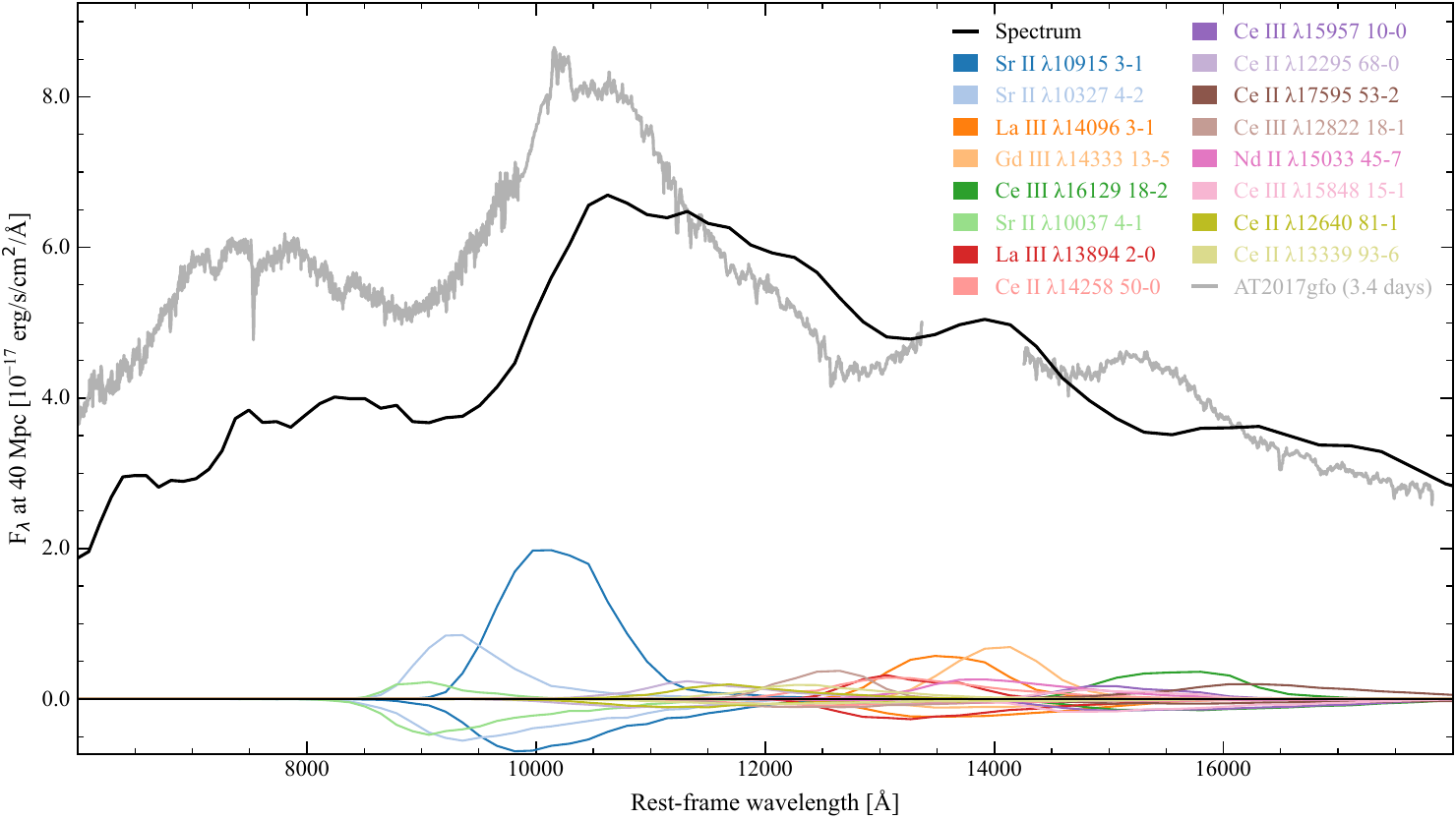}
     \end{center}
   \caption{As the top panel of Fig.~\ref{fig:spectra_ions_2p4d_3d_dyn_mass08x_veldiv2}, but decomposed by the individual transition responsible for the last emission or absorption event of each escaping packet, at 2.4 days (top) and 3.4 days (bottom).
At each epoch, the 16 transitions with the largest summed emission and absorption flux are shown.
The legend gives the ion, the rest wavelength, and the upper and lower level indices of each transition.
The level indices count the levels of each ion in order of energy, starting from 0 at the ground level, so a large index marks a level far above the ground level.}
     \label{fig:spectra_lines_2p4d_3d_dyn_mass08x_veldiv2}
\end{figure*}

\begin{table}
\caption{The nine transitions with the largest contribution to the polar spectrum of the 3D dyn 8x mass v/2 model at 2.4 days, ranked by the sum of their unrounded emission and absorption fluxes, and \ion{Sr}{ii} $\lambda$4216, which shapes the blue cut-off.}
\label{tab:toplines}
\centering
\setlength{\tabcolsep}{2.5pt}
\begin{tabular}{llrrl}
\hline\hline
Ion & $\lambda_\mathrm{rest}$ & Emission & Absorption & Feature\\
 & (\AA) & \multicolumn{2}{c}{($10^{-14}$ erg s$^{-1}$ cm$^{-2}$)} & \\
\hline
\ion{Sr}{ii}  & 10915 & 3.7 & 1.5 & NIR peak\\
\ion{Sr}{ii}  & 10327 & 1.4 & 1.4 & NIR peak\\
\ion{La}{iii} & 14096 & 1.3 & 0.76 & 13\,500~\AA\ shoulder\\
\ion{Gd}{iii} & 14333 & 1.4 & 0.38 & 13\,500~\AA\ shoulder\\
\ion{Ce}{iii} & 16129 & 1.0 & 0.53 & NIR pseudo-continuum\\
\ion{Y}{ii}   & 7882  & 1.2 & 0.34 & 7300~\AA\ peak (blue side)\\
\ion{Th}{iii} & 11382 & 1.1 & 0.41 & NIR peak (red wing)\\
\ion{Sr}{ii}  & 10037 & 0.47 & 0.98 & NIR trough\\
\ion{La}{iii} & 13894 & 0.62 & 0.74 & 13\,500~\AA\ shoulder\\
\ion{Sr}{ii}  & 4216  & 0.02 & 0.62 & blue cut-off\\
\hline
\end{tabular}
\tablefoot{NIR: near-infrared.
Emission and absorption fluxes are last-interaction quantities of the escaping packets (Section~\ref{sec:featureid}), not totals over the ejecta.
The rest wavelength $\lambda_\mathrm{rest}$ of each transition is the wavelength in standard air (the convention of the NIST Atomic Spectra Database) from that database where it contains the line, and otherwise the AD3 vacuum wavelength converted to air.
The emergent features peak several hundred \AA\ blueward of the rest wavelength because of the ejecta velocity.
Fluxes are given at a distance of 40~Mpc.
The integrated flux of the model spectrum between 3500 and 22\,000~\AA\ is $9.7\times10^{-13}$ erg s$^{-1}$ cm$^{-2}$ at the same distance.
The strongest \ion{Pr}{iii} transition, $\lambda$11406, ranks outside the top 30.
The order of transitions with similar fluxes depends on the uncertain transition probabilities.}
\end{table}

Figure~\ref{fig:spectra_lines_2p4d_3d_dyn_mass08x_veldiv2} shows the same decomposition by individual transition, with the 16 transitions that have the largest summed emission and absorption flux at each epoch (2.4 and 3.4 days), and Table~\ref{tab:toplines} lists the strongest of them at 2.4 days.
In the model, the near-infrared peak near 10\,500~\AA, the narrow peak near 7300~\AA, and the 13\,500~\AA\ shoulder are produced by single transitions or small groups of transitions that stand out above the pseudo-continuum.
This smooth pseudo-continuum is the sum of many overlapping emission and absorption contributions.
However, the strongest single transition carries only about 4~per cent of the integrated flux, and the transitions outside the top 16 supply most of the emission at every wavelength.
In their set of 3D merger models with calibrated lanthanide data, \citet{collins_exploring_2026} also attribute the near-infrared peak to the \ion{Sr}{ii} triplet and the 13\,500~\AA\ shoulder to \ion{La}{iii} and \ion{Gd}{iii}.

The near-infrared peak receives its largest contribution from the \ion{Sr}{ii} $\lambda\lambda$10037, 10327, 10915 triplet, with $\lambda$10915 alone carrying more than twice the emission of any other transition in the spectrum.
These three lines form one multiplet with common upper and lower terms, so we consider the triplet as a whole.
\citet{watson_identification_2019} attribute a P~Cygni feature in the spectra of AT2017gfo to this triplet.

The emission of \ion{Y}{ii} $\lambda$7882 peaks near 7100~\AA, and it is part of the \ion{Y}{ii} emission that contributes most to the model peak near 7300~\AA\ (Section~\ref{sec:ioncontributions}).
This transition does not form the broad optical peak itself, because it carries only $\sim$1~per cent of the integrated model flux (Table~\ref{tab:toplines}).
\citet{sneppen_discovery_2023} attribute a P~Cygni feature near 7600~\AA\ in the spectra of AT2017gfo to the \ion{Y}{ii} multiplet that contains this transition, but we do not identify the model peak with an observed feature (Section~\ref{sec:ioncontributions}).

The shoulder near 13\,500~\AA\ receives its largest contributions from the \ion{La}{iii} $\lambda\lambda$13894, 14096 pair together with \ion{Gd}{iii} $\lambda$14333.
These are the same two \ion{La}{iii} transitions that \citet{domoto_lanthanide_2022} proposed for the broad absorption feature near 12\,000--13\,000~\AA\ in AT2017gfo, and that \citet{gillanders_modelling_2024} and \citet{collins_exploring_2026} also find in their models.
Our model produces them as the two strongest \ion{La}{iii} transitions at all three epochs (the 1.4-day decomposition is not shown).
\citet{gillanders_modelling_2024} first proposed the \ion{Gd}{iii} $\lambda$14333 transition as a candidate for the observed emission feature near 14\,000~\AA, and \citet{rahmouni_revisiting_2025} proposed the same transition for the absorption feature.

The \ion{Ce}{iii} $\lambda\lambda$15848, 15957, 16129 transitions are a further case of agreement with earlier proposals.
All three rank within the top 20 at 2.4 days, and they are the three strongest \ion{Ce}{iii} lines that \citet{domoto_lanthanide_2022} proposed for the absorption feature near 14\,000~\AA.
\citet{tanaka_cerium_2023} detected them among the strongest absorption lines in the near-infrared spectrum of the chemically peculiar star HR~465.
These identifications and our AD3 data rest on the same experimentally known energy levels.

The blue cut-off below $\sim$4500~\AA\ is shaped by the \ion{Sr}{ii} $\lambda$4216 resonance line, which acts almost purely in absorption (Table~\ref{tab:toplines}).
This line is not among the 16 transitions shown and lies outside the wavelength range of Fig.~\ref{fig:spectra_lines_2p4d_3d_dyn_mass08x_veldiv2}, but no other transition below 5000~\AA\ is stronger at 1.4 or 2.4 days.
The remaining transitions in the top 16 belong to \ion{Ce}{iii}, \ion{Th}{iii}, and \ion{Y}{ii}, and they contribute to the pseudo-continuum and form no distinct peak.

Unlike the near-infrared peak, the \ion{Y}{ii} $\lambda$7882 emission, and the 13\,500~\AA\ shoulder, the two \ion{Pr}{iii} peaks do not come from transitions that rank among the strongest of the spectrum.
No \ion{Pr}{iii} transition enters the top 16, and the strongest one, $\lambda$11406, emits in the near-infrared and carries only $\sim$6~per cent of the summed emission and absorption flux of \ion{Pr}{iii} (Table~\ref{tab:toplines}).
By contrast, the four \ion{Sr}{ii} transitions in Table~\ref{tab:toplines} supply $\sim$90~per cent of the summed emission and absorption flux of \ion{Sr}{ii}.
The two \ion{Pr}{iii} peaks are therefore the combined emission of a few tens of \ion{Pr}{iii} transitions that cluster in wavelength, and about five of them supply half of this emission (not shown).
The strength of the \ion{Pr}{iii} feature rests on theoretical transition probabilities, although the wavelengths of these transitions are experimental.
We therefore treat this feature as a candidate contribution rather than an identification.

At 3.4 days (bottom panel), the \ion{Sr}{ii} triplet persists.
Five \ion{Ce}{ii} transitions ($\lambda\lambda$12295, 12640, 13339, 14258, 17595) now join the \ion{Ce}{iii} transitions in the top 16, whereas at 2.4 days none of them did.
This change is consistent with the recombination of \ion{Ce}{iii} to \ion{Ce}{ii} as the ejecta cool.

\subsection{Comparison with the 2D model}
\label{sec:2dcomparison}
The 2D model shares the mass (to within 3~per cent), the radial velocity distribution below $0.24c$, and the azimuthally averaged composition of its 3D parent by construction.
It does not contain the fastest matter in the edge regions of the 3D grid (Section~\ref{sec:ejectamodels}).
We therefore expect the differences between the two models to come mainly from the azimuthal averaging.
The deposited decay power of the 2D model is within a few per cent of that of its 3D parent, so the two models have almost the same energy budget.
The 2D calculation on the finer $50{\times}100$ grid tests the effect of the grid resolution.
Its luminosity after 0.7 days and its spectra at 1.4, 2.4, and 3.4 days agree with those of the $25{\times}50$ grid to within a few per cent (Figs~\ref{fig:lightcurves} and \ref{fig:spectra_series}).
The coarse 2D grid therefore does not cause the differences between the 2D and 3D models.

In the bolometric light curve (Fig.~\ref{fig:lightcurves}), the 2D model is a factor of $\sim$3 fainter than its 3D parent and a factor of $\sim$4--5 fainter than the observations at $\sim$1 day.
Its luminosity then declines more slowly than that of the 3D models, crossing the observations at $\sim$7 days and remaining brighter than observed thereafter.
Its spectra (Fig.~\ref{fig:spectra_series}) are fainter and redder than those of the 3D parent at every epoch.
Their colours resemble those of the mass-scaled model more than those of the velocity-compressed 3D model, although the 2D spectra are fainter than those of both 3D models.

The bottom panel of Fig.~\ref{fig:spectra_ions_2p4d_3d_dyn_mass08x_veldiv2} shows the ion decomposition of the 2D spectrum at 2.4 days.
The same ions contribute, but the total emitted flux is less than half that of the 3D model and the spectrum is redder.
The \ion{Y}{ii} (optical) and \ion{Sr}{ii} (near-infrared) contributions that shape the 3D spectrum are strongly suppressed, and the \ion{Th}{iii} emission moves from the red wing of the near-infrared peak to beyond 17\,000~\AA.
The comparatively unblended \ion{Sr}{ii} feature near 10\,500~\AA, whose position depends on the expansion velocity (Section~\ref{sec:ioncontributions}), is weakened and shifted blueward (peaking near 9300~\AA), and it no longer stands out against the broad \ion{Ce}{ii} emission.
Azimuthal averaging therefore also changes the individual line contributions and their diagnostic value.

 \begin{figure}
     \begin{center}
     \includegraphics[width=\hsize]{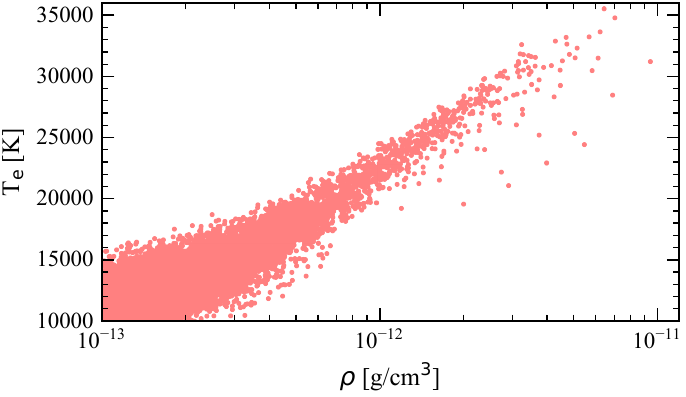}
     \includegraphics[width=\hsize]{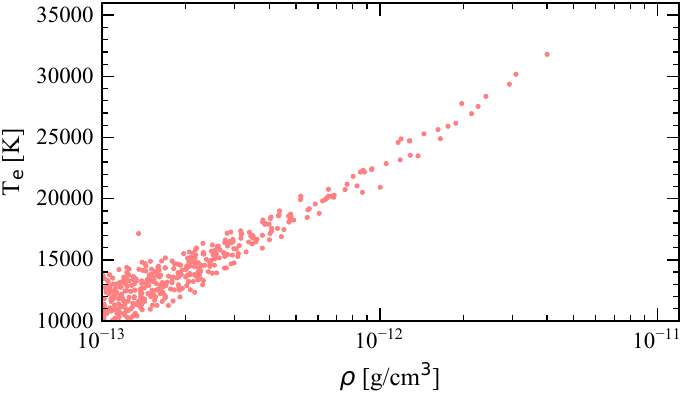}
     \end{center}
   \caption{Temperature versus mass density for the cells of the dyn 8x mass v/2 model in 3D (top) and in 2D on the $25{\times}50$ grid (bottom), for timesteps spanning 1.2--1.7 days.
Each point represents one cell at one sampled timestep.
The panels show only the cells with $\rho > 10^{-13}\,\mathrm{g\,cm^{-3}}$ and $T_e > 10\,000$~K.
The 3D model reaches about twice the maximum density of the azimuthally averaged 2D model and has a wider scatter in $T_e$ at a given density.
Each point of the 2D model is a ring that holds more mass than a cell of the 3D model.}
     \label{fig:estimators_3d_dyn_mass08x_veldiv2}
 \end{figure}

The azimuthal averaging mixes the composition, which leaves only a few per cent of the mass of the 2D model in lanthanide-poor regions (Fig.~\ref{fig:xlanvelocity}; Section~\ref{sec:ejectamodels}).
Figure~\ref{fig:estimators_3d_dyn_mass08x_veldiv2} shows that the averaging also narrows the range of thermodynamic conditions.
At $\sim$1.4 days, the cells of the 3D model span a broad range of densities and temperatures, with a wide scatter in $T_e$ among cells of similar density.
A tail of dense clumps (reaching $\rho \sim 10^{-11}\,\mathrm{g\,cm^{-3}}$) is heated to $T_e \gtrsim 25\,000$~K, hotter than most of the cells in the figure, which are at $10\,000$--$20\,000$~K.
The averaging lowers the maximum density by a factor of about two and reduces the scatter in $T_e$ at a given density.
The mixed composition and the narrower range of conditions are each consistent with the lower early luminosity of the 2D model relative to the 3D model and AT2017gfo (Fig.~\ref{fig:lightcurves}; Section~\ref{sec:3dstructure}).

\section{Results with opacity and thermal-redistribution approximations}
\label{sec:resultsopac}

\begin{figure}
     \begin{center}
     \includegraphics[width=\hsize]{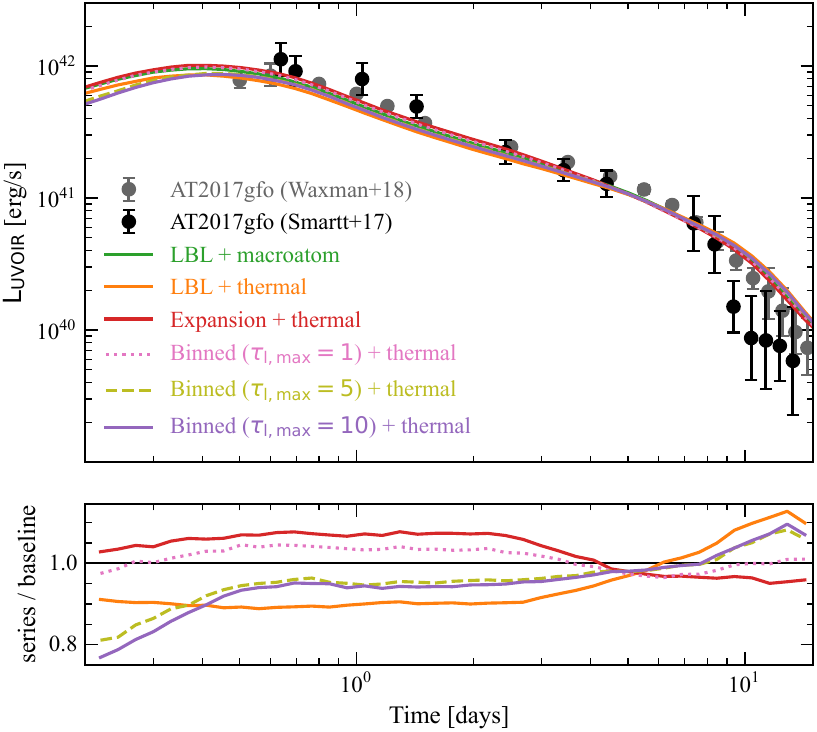}
     \end{center}
   \caption{Isotropic-equivalent bolometric luminosity (ultraviolet--optical--infrared power) in the polar direction for the 3D dyn 8x mass v/2 model computed with the line-by-line macroatom (LBL + macroatom), line-by-line thermal-redistribution (LBL + thermal), expansion-opacity (Expansion + thermal), and capped line-binned (Binned ($\tau_\mathrm{l,max}=1$) + thermal, Binned ($\tau_\mathrm{l,max}=5$) + thermal, and Binned ($\tau_\mathrm{l,max}=10$) + thermal) treatments, compared with bolometric estimates for AT2017gfo \citep{smartt_kilonova_2017,waxman_constraints_2018}.
The lower panel shows the ratio of each model to the LBL + macroatom model.}
     \label{fig:lightcurves_opacchoice}
\end{figure}

\begin{figure*}
     \begin{center}
     \includegraphics[width=0.84\hsize]{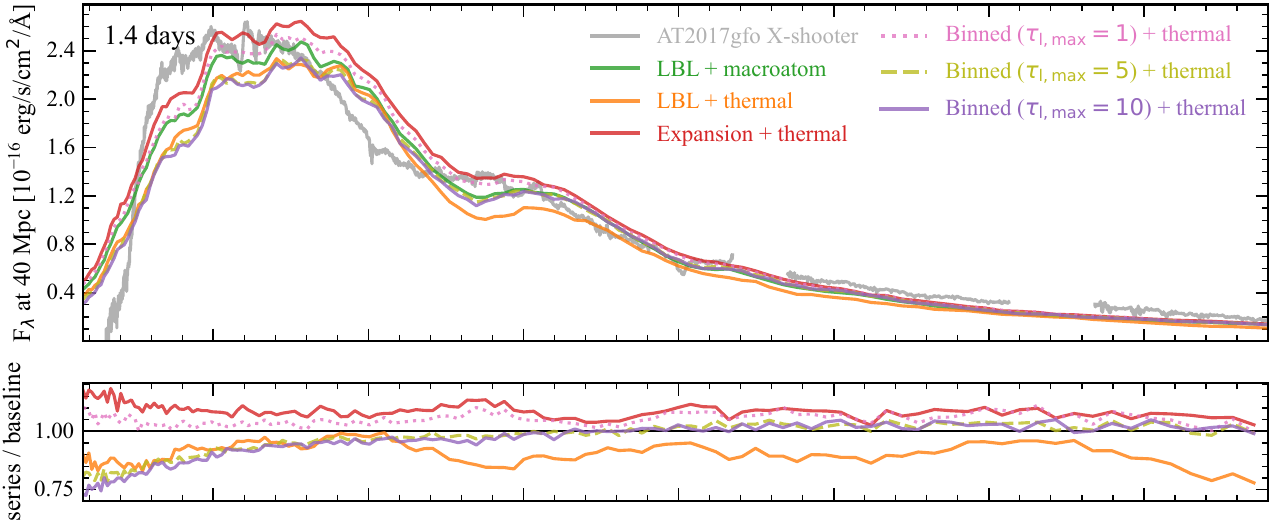}
     \includegraphics[width=0.84\hsize]{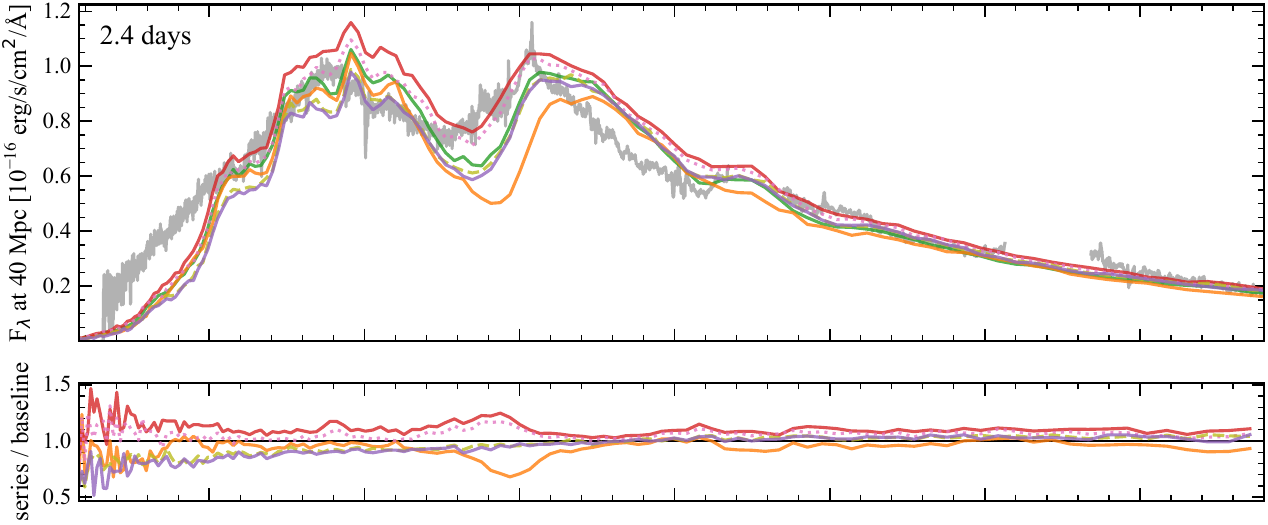}
     \includegraphics[width=0.84\hsize]{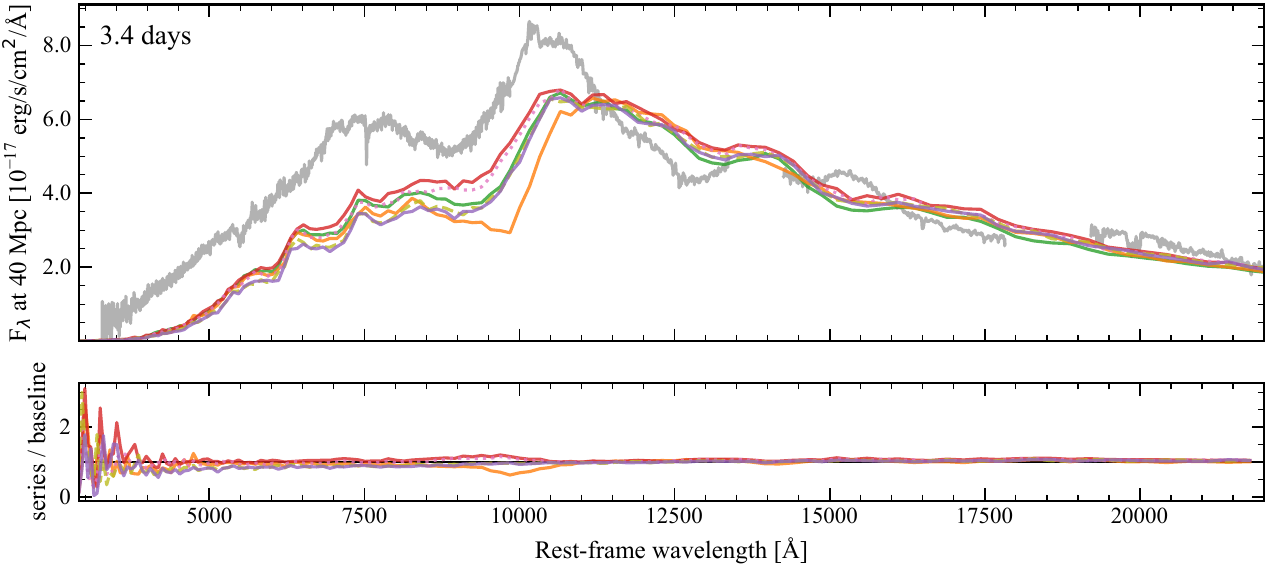}
     \end{center}
   \caption{Spectra in the polar direction at 1.4 days (top), 2.4 days (middle), and 3.4 days (bottom) for the 3D dyn 8x mass v/2 model with the line-by-line macroatom (LBL + macroatom), line-by-line thermal-redistribution (LBL + thermal), expansion-opacity (Expansion + thermal), and capped line-binned (Binned ($\tau_\mathrm{l,max}=1$) + thermal, Binned ($\tau_\mathrm{l,max}=5$) + thermal, and Binned ($\tau_\mathrm{l,max}=10$) + thermal) treatments, compared with the observed AT2017gfo spectra \citep{pian_spectroscopic_2017,smartt_kilonova_2017} at a distance of 40~Mpc.
The lower panel of each epoch shows the ratio of each model spectrum to the LBL + macroatom spectrum.}
     \label{fig:opacchoice_spectra_series}
\end{figure*}

Here we compare the six configurations of the velocity-compressed 3D model, which differ in the treatment of line opacity and emission (Section~\ref{sec:methodsemission}).
Comparing LBL + macroatom with LBL + thermal isolates the effect of the emission treatment at fixed line-by-line opacities, apart from the additional self-absorption of the LBL + thermal configuration (Section~\ref{sec:methodsemission}).
Comparing LBL + thermal with the four binned-opacity configurations then approximately isolates the effect of binning the lines.
All five use thermal redistribution, but not all of them use the same binned line opacity in the Kirchhoff emissivity (Section~\ref{sec:methodsemission}).
In this section, `the saturated binned models' means the Expansion + thermal and Binned ($\tau_\mathrm{l,max}=1$) + thermal models, in which a strong line saturates at an optical depth of one.

Figure~\ref{fig:lightcurves_opacchoice} shows the UVOIR luminosity evolution for the six configurations.
We find that replacing the macroatom method with full thermal redistribution makes the model $\sim$10~per cent fainter before $\sim$3 days.
The two line-by-line curves cross at $\sim$6 days, and the LBL + thermal model is $\sim$10~per cent brighter after $\sim$10 days.
The emission treatment does not change the total energy budget, but it sets the wavelengths at which the energy is re-emitted and thus the opacity that the re-emitted photons encounter.
It therefore controls when the energy escapes.
Thermal re-emission at the local electron temperature puts more energy at blue and ultraviolet wavelengths than fluorescence does.
At the ejecta temperatures of the first days (Fig.~\ref{fig:estimators_3d_dyn_mass08x_veldiv2}), the Planck function in the Kirchhoff emissivity peaks in the ultraviolet, and the line opacity that weights this function is largest at short wavelengths (Fig.~\ref{fig:expansionopacitycomparison}).
We therefore attribute the early deficit and the late excess of the LBL + thermal model to a longer trapping of the radiation.

Relative to LBL + thermal, binning the line opacities raises the luminosity.
The two saturated binned models agree with each other to within a few per cent and are $\sim$20~per cent brighter than the LBL + thermal model before $\sim$2.5 days.
Part of this difference comes from the additional self-absorption of the LBL + thermal configuration (Section~\ref{sec:methodsemission}).
Relative to the macroatom model, the saturated binned models are less than 10~per cent brighter between 0.5 and 2 days.
The line-binned models with $\tau_\mathrm{l,max} = 5$ and 10, which we call the converged line-binned models (see below), are instead less than 10~per cent fainter than the macroatom model between 0.5 and 4 days.
The saturation of the strong lines thus raises the luminosity by $\sim$10~per cent relative to the converged line-binned models.
As this excess offsets the deficit from the thermal redistribution, the saturated binned models are closer to the macroatom model than the LBL + thermal model is.
At $\sim$5 days, the LBL + macroatom, LBL + thermal, and saturated binned models agree to within a few per cent.

Figure~\ref{fig:opacchoice_spectra_series} shows the polar spectra at 1.4, 2.4, and 3.4 days for the six configurations.
At 1.4 days, the spectra of the LBL + macroatom, LBL + thermal, and saturated binned models have the same shape and differ mainly in flux level.
At all three epochs, the saturated binned models exceed the macroatom flux by up to $\sim$20~per cent, with the largest excess in the trough between the optical and near-infrared peaks at 2.4 and 3.4 days.
The size and location of the excess change with epoch.
The LBL + thermal model is instead fainter than the macroatom model, by up to $\sim$10~per cent in the optical at 1.4 days and by 20--30~per cent in the trough at 2.4 days.
Redward of $\sim$12\,000~\AA, the spectra of these four models agree to within $\sim$10~per cent at 2.4 and 3.4 days.

The Expansion + thermal and Binned ($\tau_\mathrm{l,max}=1$) + thermal spectra track each other closely (within about 10~per cent) at all three epochs, with the expansion-opacity spectrum slightly brighter.
This ordering is expected because the line-binned opacity is never smaller than the expansion opacity for the same level populations (Section~\ref{sec:cappedopac}).
Because the two opacities differ by only $\sim$10--20~per cent (Fig.~\ref{fig:expansionopacitycomparison}), replacing $1-e^{-\tau_i}$ with $\min(\tau_i, 1)$ has little additional effect on the emergent spectra.

We attribute the flux excess of the saturated binned models to the packets that cross bins with optically thick lines without an interaction (Section~\ref{sec:binTransport}; Table~\ref{tab:binprob}).
With the expansion opacity and with $\tau_\mathrm{l,max} = 1$, a bin that contains one strong line has a probability of no interaction of $P_0 = e^{-1} \approx 0.37$, although almost every packet interacts with the line in the line-by-line treatment.
This equal probability explains why the two saturated binned models agree so closely with each other.
In bins with many optically thick lines, almost every packet interacts whether or not the strong lines saturate, which can explain why the saturation raises the bolometric luminosity by only $\sim$10~per cent.
Even where the bins hold few optically thick lines, a packet that crosses one bin with a strong line without an interaction can meet another bin with a strong line before it escapes.
Binning can then shift the point of absorption by a few bins rather than let the packet escape.

The excess of the Expansion + thermal model is about 10~per cent at most wavelengths but reaches about 20~per cent near 9000--10\,000~\AA\ at 2.4 days and between 8500 and 10\,000~\AA\ at 3.4 days.
These are wavelength regions where the ejecta are partly transparent in the line-by-line calculation, and where a bin therefore holds only a few optically thick lines.
Below $\sim$5000~\AA, where the line blanketing is strongest and the flux is small, the Expansion + thermal model is nevertheless 10--20~per cent brighter at 2.4 days, so the excess is not limited to the partly transparent regions.
At 3.4 days, Monte Carlo noise dominates the ratios in this region.

To test the convergence towards the uncapped line-binned opacity, the Binned ($\tau_\mathrm{l,max}=5$) + thermal and Binned ($\tau_\mathrm{l,max}=10$) + thermal models use higher caps, which reduce $P_0$ for a bin with one strong line to $e^{-5} \approx 0.007$ and $e^{-10} \approx 4.5\times10^{-5}$.
With these caps, the flux excess of the saturated binned models disappears.
Between 0.5 and 4 days, the two models are less than 10~per cent fainter than the macroatom model, and they agree with it to within a few per cent at 5--8 days (Fig.~\ref{fig:lightcurves_opacchoice}).
Relative to LBL + thermal, binning with these caps raises the luminosity between 0.5 and 2.5 days by less than 10~per cent, compared with $\sim$20~per cent for the saturated binned models.
At all three epochs, the spectra of the converged line-binned models agree with the macroatom spectrum to within 10~per cent redward of $\sim$7500~\AA\ (Fig.~\ref{fig:opacchoice_spectra_series}), while they are 10--20~per cent fainter than the macroatom spectrum blueward of $\sim$5500~\AA.
In the blue, they are similar to or fainter than the LBL + thermal spectrum.

The luminosities and spectra with $\tau_\mathrm{l,max} = 5$ and 10 differ from each other by only a few per cent, so the results converge as the cap increases.
This convergence suggests that the uncapped line-binned opacity, which is too costly to run in ARTIS (Section~\ref{sec:cappedopac}), would give similar results.
The cap needed for convergence depends on the distribution of the line optical depths, so other ejecta models or bin widths may need a different cap.

The LBL + thermal model has a deeper trough near 9500~\AA\ than all other models, and absorption by \ion{Sr}{ii} $\lambda$10037 causes this trough (Table~\ref{tab:toplines}).
Because the converged line-binned models also use thermal redistribution and do not have the deeper trough, we attribute the greater depth of this trough to the additional self-absorption of the LBL + thermal configuration (Section~\ref{sec:methodsemission}) and not to thermal redistribution.

\section{Discussion}
\label{sec:discussion}
The synthetic spectra depend on four ingredients: atomic wavelengths and transition probabilities, the ejecta mass--velocity distribution, the multidimensional structure of the ejecta, and the treatment of line opacity and emission.
Our model set varies the last three.
Within the present model set, the mass--velocity distribution and the reduction from three dimensions to two change the spectra more than the line treatments do.
In our velocity-compressed model, the line treatments change the emergent flux by up to 30~per cent, and the size of the change depends on epoch and wavelength.

The AD3 atomic data place the contributions of heavy ions at measured wavelengths where experimental energy levels exist, so the gaps and peaks of the line blend can be compared with the observations.
Many overlapping emission and absorption contributions form the pseudo-continuum, and fitting isolated line profiles (e.g., P~Cygni models) to the more blended features can therefore give misleading velocities and abundances unless the surrounding blend is modelled at the same time.

\subsection{Sensitivity to the ejecta mass--velocity distribution}
\label{sec:discmassvel}
The velocity compression prevents the premature reddening of the models with the original velocities (Section~\ref{sec:resultsscaled}).
Halving the velocities at fixed mass raises the densities by a factor of eight, and every Sobolev optical depth would then be eight times larger at a given epoch if the temperature and ionisation state were unchanged.
Although the temperature and ionisation also respond to the higher density, the net effect is that the ejecta remain optically thick for longer and trap the radiation.

In homologous expansion, the velocity of the photosphere $v_{\mathrm{ph}}$ also sets its radius, $R_{\mathrm{ph}} = v_{\mathrm{ph}} t$.
The lower expansion velocities of the velocity-compressed model give it a smaller radiating surface, and after $\sim$0.4 days it is also more luminous than the mass-scaled model (Fig.~\ref{fig:lightcurves}).
Both differences raise the effective temperature ($T_{\mathrm{eff}} \propto L^{1/4} (v_{\mathrm{ph}} t)^{-1/2}$, where $L$ is the luminosity), which gives bluer colours.
The same compression also raises the fraction of the decay energy that is deposited (Section~\ref{sec:lightcurves}), so the velocity distribution affects both the energy budget and the colours.

The mass-scaled model shows that the bolometric light curve alone constrains the mass--velocity distribution only weakly.
This model is within $\sim$30~per cent of the observed luminosity between 1.5 and 5 days, yet at 1.4 days its spectral energy distribution peaks at almost twice the observed wavelength.
Agreement with the bolometric light curve is therefore not a sufficient test of an ejecta model, and a fit to the light curve alone can hide large errors in the mass--velocity distribution.

\subsection{The role of 3D ejecta structure}
\label{sec:3dstructure}
A 2D version of the scaled 3D ejecta is a factor of $\sim$3 fainter than its 3D parent and a factor of $\sim$4--5 fainter than AT2017gfo at $\sim$1 day, and it has redder spectra than its 3D parent.
It nevertheless shares by construction the mass, the radial velocity distribution, and the azimuthally averaged composition of its 3D parent, and it has almost the same deposited decay power (Section~\ref{sec:2dcomparison}).
Two effects of the azimuthal averaging can cause the fainter and redder emission.
First, the averaging mixes the composition, which leaves only a few per cent of the mass of the 2D model in lanthanide-poor regions (Fig.~\ref{fig:xlanvelocity}; Section~\ref{sec:ejectamodels}).
A lanthanide mass fraction of only $\sim$$10^{-3}$ already raises the opacity by orders of magnitude \citep{kasen_opacities_2013}, so the mixing raises the opacity of most of the matter that is lanthanide-poor in the 3D model.
Second, the averaging smooths the density, which reduces the contrast between the dense clumps and the low-density channels between them (Fig.~\ref{fig:estimators_3d_dyn_mass08x_veldiv2}).
Since radiation escapes a clumpy medium mainly through its low-density regions, smoothed ejecta have a higher effective optical depth than clumpy ejecta with the same mean density.
The smoothing lowers the mass-weighted mean density most in the outer ejecta (Section~\ref{sec:ejectamodels}).

Our comparison does not separate the two effects (Section~\ref{sec:limitations}).
The mixing and the smoothing should each trap the radiation for longer.
Longer trapping is consistent with the 2D model being a factor of $\sim$3 fainter than its 3D parent at 1 day and then declining more slowly as the trapped energy is released.
At the same expansion velocities, this lower early luminosity gives a lower effective temperature (Section~\ref{sec:discmassvel}) and a redder spectrum.

Inferences of ejecta properties from spherically or azimuthally averaged versions of intrinsically 3D ejecta should therefore be treated with caution, because the averaged models can be significantly fainter at early times and redder than their 3D parents.
\citet{shingles_selfconsistent_2023} find the corresponding result for spherical averages: 1D versions of the 3D ejecta give different luminosities and spectra, with less flux below $\sim$8000~\AA.
The high degree of sphericity inferred for the AT2017gfo photosphere from the \ion{Sr}{ii} P~Cygni feature and the blackbody normalisation \citep{sneppen_spherical_2023} may not conflict with asymmetric ejecta.
\citet{collins_inferring_2024} show that a 3D model with asymmetric ejecta can appear more spherical in such measurements than the ejecta are.

\subsection{Implications for inferred ejecta properties}
Our velocity compression is a substitute for the slower post-merger ejecta that the model lacks, and it is not the result of a hydrodynamic simulation.
The comparison of the mass-scaled model with the velocity-compressed model shows that the mass--velocity distribution strongly affects the early spectra.
The same comparison does not show how the mass divides between the dynamical and the post-merger ejecta.
The distribution that approximates the early spectra has a mass of $\sim$0.04~\Msun, with half of this mass below $v_r \sim 0.09c$ (Fig.~\ref{fig:density}), and approaches the mass and velocity range predicted for viscously driven outflows from the post-merger disc.
The end-to-end models of \citet{just_endtoend_2023} eject up to $\sim$0.06~\Msun from the black-hole disc at mean velocities of $\sim$0.03--$0.06c$ (see also, e.g., \citealt{fernandez_delayed_2013,siegel_threedimensional_2018,fujibayashi_viscous_2020} for earlier models of such outflows).
Other post-merger outflows that may partly reproduce the velocity-compressed model are neutrino-driven winds \citep[e.g.,][]{perego_neutrinodriven_2014,fujibayashi_postmerger_2020,just_endtoend_2023}, spiral-wave winds \citep{nedora_spiralwave_2019}, and magnetically driven winds \citep[e.g.,][]{combi_jets_2023,curtis_magnetized_2024,kiuchi_firstprinciples_2026}.
The early spectra may thus be consistent with a mass--velocity distribution like that of the post-merger ejecta, even though our model has the composition of the dynamical ejecta.

\subsection{Implications for a separate lanthanide-poor component}
The velocity-compressed model, whose mean lanthanide mass fraction is 0.023, reproduces the optical flux of AT2017gfo redward of 5500~\AA\ at 1.4 and 2.4 days.
Between 4500 and 5500~\AA, it gives most of the observed flux at both epochs.
Between 3500 and 4500~\AA, it gives most of the observed flux at 1.4 days and less than half at 2.4 days, so the blue deficit grows with time.
Between 5500 and 15\,000~\AA, and apart from the trough near 9500~\AA, the model falls short of the observations only at 3.4 days.
One ejecta model with a single composition distribution thus approximates the spectra of AT2017gfo at 1.4 and 2.4 days, provided that it keeps its 3D structure and has a mass--velocity distribution like that of the velocity-compressed model.

This composition is not uniformly lanthanide-rich, as about a quarter of the mass is in regions with $X_\mathrm{lan} < 10^{-3}$ that are interleaved in three dimensions with the lanthanide-rich matter (Section~\ref{sec:ejectamodels}).
The model gives this blue and optical flux only with its 3D structure and its mass--velocity distribution.
The azimuthally averaged 2D version of the same ejecta, in which only a few per cent of the mass remains lanthanide-poor, is too red (Section~\ref{sec:2dcomparison}), and the 1D averages of \citet{shingles_selfconsistent_2023} are likewise redder than their 3D ejecta.
The models with the original velocities are also too red (Section~\ref{sec:resultsscaled}).

The multicomponent interpretations of the early emission \citep[e.g.,][]{kasen_origin_2017,perego_2017gfo_2017,villar_combined_2017} describe the ejecta as separate components, each with its own mass, velocity, and opacity or composition.
These interpretations rest on 1D or parameterised ejecta, whose lanthanide-rich components are too red to supply the early optical flux.
The emission of AT2017gfo redward of 4500~\AA\ at 1.4--2.4 days therefore does not require a separate lanthanide-poor ejecta component \citep[in contrast to the interpretation of, e.g.,][]{arcavi_optical_2017}.
Within our model family, this emission does require more mass at lower velocities than the dynamical ejecta contain, but this additional mass does not need a composition that differs from that of the dynamical ejecta.
The model needs no lanthanide-poor matter beyond the quarter of the mass that this composition contains, provided that the ejecta keep their 3D structure.

The multicomponent interpretations invoke the separate lanthanide-poor component mainly for the blue emission before $\sim$1.5 days \citep[see][for a compilation of the spectra of this phase]{sneppen_emergence_2024}.
Most of this phase precedes our earliest spectral epoch at 1.4 days and is outside our spectral comparison, and the bolometric light curve is the only observable that we compare before 1.4 days.
Between 0.8 and 1.5 days, the model is fainter than both observational estimates, and a correction of the snapshot thermal energy would make it fainter (Section~\ref{sec:lightcurves}).
The case for a separate lanthanide-poor component in this earlier phase should be tested with multidimensional models whose composition comes from merger simulations.

The conclusion that the optical emission at 1.4--2.4 days needs no separate lanthanide-poor component does not depend on the opacity treatment.
All six configurations of Section~\ref{sec:resultsopac} give the integrated flux between 4000 and 7500~\AA\ at 1.4 and 2.4 days to within $\sim$10~per cent of the line-by-line macroatom result, so the line treatments change the flux evolution quantitatively and not qualitatively.

The treatments are nevertheless relevant to other studies because many published kilonova simulations use expansion opacities with thermal redistribution \citep[e.g.,][]{tanaka_radiative_2013,kasen_origin_2017,kawaguchi_lowmass_2021}.
In our comparisons, these approximations change the bolometric light curve by $\lesssim$10~per cent and change the spectrum little redward of $\sim$12\,000~\AA.
They overestimate the flux relative to the line-by-line macroatom calculation by 10--20~per cent, with the largest excess in the trough near 9500~\AA\ (Section~\ref{sec:resultsopac}).
We attribute the excess to the saturation of the strong lines in the expansion opacity (Section~\ref{sec:binTransport}), and not to the thermal redistribution or to the binning itself.
The line-by-line calculation with thermal redistribution is fainter than the macroatom result at all three epochs, and the converged line-binned calculations show no excess.

Because we attribute the excess to the saturation of strong lines, which matters most where a bin contains a few optically thick lines, we expect the bias to depend on the lanthanide mass fraction and the density of the ejecta.
Lanthanide mass fractions and component masses inferred with expansion opacities can therefore be affected by a systematic flux bias, which is 10--20~per cent in our model and depends on epoch and wavelength.
The bias makes a lanthanide-rich composition appear brighter in the optical than it is, and is therefore unlikely to be the reason that models with expansion opacities need a lanthanide-poor component.
This bias does not apply to the line-binned opacities of \citet{fontes_linebinned_2020}, which do not saturate for strong lines and which a code with a diffusion treatment can use \citep[e.g.,][]{even_composition_2020,korobkin_axisymmetric_2021,wollaeger_broad_2021}.
ARTIS cannot use them directly, but our capped calculations converge between $\tau_\mathrm{l,max} = 5$ and 10, and they show no flux excess.
Instead, the flux blueward of $\sim$5500~\AA\ is below the macroatom result (Section~\ref{sec:resultsopac}).

\subsection{Limitations}
\label{sec:limitations}
Residual wavelength offsets and uncertain transition strengths in the atomic data propagate directly into the feature assignments.
This matters most for \ion{Th}{iii}, which supplies more than half of the 16 strongest transitions at 1.4 days (not shown), four at 2.4 days, and none at 3.4 days (Fig.~\ref{fig:spectra_lines_2p4d_3d_dyn_mass08x_veldiv2}).
These contributions rest on the theoretical atomic data of \citet{kitoviene_theoretical_2025}, which are not calibrated to measured energy levels.
The wavelengths of these data are uncertain by several hundred \AA\ in the near-infrared, so we do not treat the \ion{Th}{iii} contributions as identifications (Section~\ref{sec:featureid}).

The \ion{Th}{iii} population is also an upper limit because \ion{Th}{iii} is the only Th ionisation stage in AD3, so all Th is doubly ionised in every cell at all times.
La and Ce have four stages in AD3, and their second and third ionisation energies are within $\sim$2~eV of those of Th \citep{kramida_nist_2024}.
In the velocity-compressed model as a whole, about a fifth of the La is doubly ionised at 1.4 days and about half at 2.4 and 3.4 days.
Since Th is harder to ionise to the doubly ionised stage than La and Ce, and easier to ionise beyond it, the doubly ionised fraction of Th would be lower than that of La.
The \ion{Th}{iii} populations are thus too high by a factor of two or more at 2.4 days, so the \ion{Th}{iii} fluxes are also too high.
At 2.4 days, the \ion{Th}{iii} emission lies mainly between 10\,500 and 12\,500~\AA\ (Section~\ref{sec:featureid}), where the model exceeds the observed flux (Fig.~\ref{fig:spectra_lines_2p4d_3d_dyn_mass08x_veldiv2}), so this overestimate may contribute to the broad red side of the near-infrared peak.
The Th abundance is also uncertain because the actinide yields of the network depend on the fission treatment and on the nuclear mass model far from stability \citep[e.g.,][]{barnes_kilonovae_2021}.

Our scaled models are not the result of a self-consistent evolution from merger to homologous expansion, and degeneracies between total mass, velocity distribution, and geometry remain.
Our velocity scaling stands in for the slower post-merger ejecta, such as the neutrino-driven and viscously driven outflows from the remnant and its disc \citep[e.g.,][]{fernandez_delayed_2013,perego_neutrinodriven_2014,just_endtoend_2023}, which long-term simulations can follow to homologous expansion \citep[e.g.,][]{kawaguchi_lowmass_2021}.
The scalings keep the elemental mass fractions, so the electron fraction and the abundances of the scaled models remain those of the dynamical ejecta, whereas post-merger ejecta are expected to be poorer in lanthanides (Section~\ref{sec:ejectamodels}).
We test only one mass factor and one velocity factor, and the velocity scaling is uniform, so it also slows the fastest ejecta.

The 2D comparison shows the effect of the azimuthal averaging as a whole, but it cannot separate the mixing of the composition from the smoothing of the density, because the averaging changes both together.
The temperature and the ionisation respond to both changes.
The degree of clumping may also depend on the hydrodynamic method, and we have not tested how much the SPH scheme of the merger simulation affects the clumps.

Our calculations assume LTE for the ionisation and excitation state of the gas.
LTE is expected to hold best in the dense, optically thick ejecta of the first few days, but departures from LTE will grow as the ejecta expand and become less dense \citep[e.g.,][]{hotokezaka_nebular_2021,pognan_nlte_2022,pognan_nlte_2023}.
\citet{brethauer_nonthermal_2026} find that non-thermal ionisation by the decay electrons raises the ionisation state above its LTE value already within the first two days, mainly in the fast, low-density ejecta, which reduces the line blanketing in the optical.
The atomic dataset also limits the ionisation because AD3 ends at the triply ionised stage for most elements (Section~\ref{sec:atomicdata}).
About three quarters of the La and Ce in the velocity-compressed model are in this highest stage at 1.4 days and about a third at 2.4 days, mainly in the dense inner ejecta.
Above $0.15c$, the share is about a fifth at 1.4 days and negligible at 2.4 days.
The true ionisation can therefore be higher than in the model in the hottest regions during the first two days, and the model does not include the opacity of the higher stages.
The limit of the dataset is therefore a larger source of uncertainty at 1.4 days than at 2.4 days.

\section{Conclusions}
\label{sec:conclusions}
We have computed radiative transfer models for the dynamical ejecta of a neutron star merger simulation with the AD3 atomic dataset, which includes wavelength-calibrated lanthanide data.
The model family comprises the unscaled 3D ejecta with 0.005~\Msun, a mass-scaled version with 0.04~\Msun, the velocity-compressed version with the same mass and halved velocities, and 2D azimuthal averages of the velocity-compressed model on two grids.
For the velocity-compressed model, we compared six treatments of line opacity and emission: line-by-line opacities with the macroatom method or thermal redistribution, the expansion opacity, and capped line-binned opacities with three caps.
We compared the light curves and the spectra at 1.4, 2.4, and 3.4 days with AT2017gfo and traced the model features to individual transitions.
Within this model family, we conclude the following:
\begin{itemize}
\item Mass scaling together with velocity compression approximates the light curve and early spectral evolution of AT2017gfo remarkably well.
To our knowledge, this is the first multidimensional kilonova model based on hydrodynamic simulations and calibrated atomic data that reproduces both the luminosity and main spectral features of AT2017gfo at early times.
The model has a flux deficit at 3.4 days, and below 4500~\AA\ it gives less than half of the observed flux at 2.4 days.
Of our 3D models, only the model with halved velocities approaches the observed optical flux at 1.4 and 2.4 days.
This configuration, a mass of $\sim$0.04~\Msun with half of this mass below $v_r \sim 0.09c$, suggests that a significant contribution to the emission of AT2017gfo may be associated with post-merger ejecta (e.g., from viscous black-hole disc winds) that are slower and more massive than dynamical ejecta, while the total (dynamical plus post-merger) ejecta could have a comparable composition to the dynamical ejecta.
Agreement with the bolometric light curve alone is not a sufficient test of an ejecta model, because the mass-scaled model is close to the observed luminosity but has much redder spectra.
The spectra thus constrain the velocity distribution of the ejecta, which the bolometric light curve constrains only weakly.
\item The velocity-compressed model produces most of the observed flux between 4500 and 5500~\AA\ at 1.4 and 2.4 days with all six opacity and emission treatments.
This emission therefore does not require a separate lanthanide-poor component \citep[in contrast to the interpretation of, e.g.,][]{arcavi_optical_2017}.
The additional mass at lower velocities that this emission requires in our models can have the relatively lanthanide-rich composition of the dynamical ejecta (mean lanthanide mass fraction of 0.023).
The velocity-compressed model produces this emission only with its 3D structure and its halved velocities.
In this structure, about a quarter of the mass is in lanthanide-poor regions that are interleaved with the lanthanide-rich matter.
The halved velocities give a smaller radiating surface and a higher effective temperature.
The blue emission before 1.4 days is outside our spectral comparison.
\item An azimuthally averaged 2D version of the scaled 3D ejecta is a factor of $\sim$3 fainter than its 3D parent and a factor of $\sim$4--5 fainter than AT2017gfo at $\sim$1 day, and it has redder spectra than its 3D parent.
We attribute these differences to two changes that our comparison does not separate.
The averaging mixes lanthanides into the lanthanide-poor matter, which raises the opacity of this matter.
The averaging also reduces the density contrast between the clumps and the low-density channels.
Both changes should trap the radiation for longer and so delay and redden the emission.
Inferences of ejecta properties from averaged versions of intrinsically 3D ejecta should therefore be treated with caution.
\item The AD3 dataset places the lanthanide features at calibrated wavelengths where experimental energy levels are available \citep{flors_calibrated_2026}, which is needed to link features in kilonova spectra to individual transitions and thus to identify the elements that produce them (Section~\ref{sec:atomicdata}).
With these data and the decomposition of the escaping packets by their last interaction, we find the transitions that contribute most to the main features of the model spectrum.
The near-infrared peak and its blueward trough receive their largest contributions from the \ion{Sr}{ii} $\lambda\lambda$10037, 10327, 10915 triplet.
The comparatively unblended \ion{Sr}{ii} feature of the velocity-compressed model peaks close to the observed feature near 10\,500~\AA, but it is broader than observed.
The emission of \ion{Y}{ii} $\lambda$7882 peaks near 7100~\AA\ and contributes to the model peak near 7300~\AA.
\ion{La}{iii} $\lambda\lambda$13894, 14096 and \ion{Gd}{iii} $\lambda$14333 contribute most to the 13\,500~\AA\ shoulder, and the \ion{Sr}{ii} $\lambda$4216 resonance line shapes the blue cut-off.
We also find a candidate contribution from \ion{Pr}{iii} near 6700 and 7300~\AA\ that arises from a few tens of transitions.
The observed spectrum at 2.4 days has one broad maximum between these wavelengths, so we do not identify this contribution with an observed feature.
Many weaker transitions contribute to each feature, and we therefore caution against fitting isolated line profiles to the more blended features of kilonova spectra.
\item Replacing the macroatom method with full thermal redistribution at fixed line-by-line opacities changes the bolometric light curve by $\sim$10~per cent and lowers the optical flux.
The expansion opacity also changes the bolometric light curve by $\lesssim$10~per cent.
In our model, however, it raises the emergent flux relative to the line-by-line macroatom calculation by 10--20~per cent, most strongly in the trough near 9500~\AA, and the size and location of the excess change with epoch.
We attribute the excess to the saturation of strong lines, because an optically thick line contributes an optical depth of only one to its bin in the expansion opacity, whereas the line absorbs almost every packet in the line-by-line treatment.
The line-binned opacity does not saturate in this way, and our capped line-binned calculations, which converge between $\tau_\mathrm{l,max} = 5$ and 10, show no flux excess.
Lanthanide mass fractions and component masses inferred with expansion-opacity treatments can therefore be systematically biased by this excess.
\end{itemize}
The 3D dynamical ejecta of a merger simulation, with their composition unchanged, can thus approximate the spectra of AT2017gfo at 1.4 and 2.4 days, provided that their mass--velocity distribution approaches that predicted for post-merger outflows.
This configuration is one ejecta model that approximates these spectra.
We have not shown that it is the only one.
The degeneracies between mass, velocity distribution, and geometry remain, and Section~\ref{sec:limitations} describes these degeneracies and the approximations of this study.
This work is a step towards fully self-consistent kilonova models that reproduce both the light curve and the spectra of AT2017gfo.
Since our model contains only the dynamical ejecta, future work is needed to determine how the spectra change with the complete 3D ejecta of a merger, as well as with departures from LTE and calibrated atomic data for the actinides.

\section*{Acknowledgements}
LJS acknowledges Gururaj Wagle for helpful discussions about line-binned opacities.

LJS, OJ, VV, FMcN, SAS and AB are funded by the European Union (ERC, HEAVYMETAL, 101071865).
Views and opinions expressed are, however, those of the authors only and do not necessarily reflect those of the European Union or the European Research Council.
Neither the European Union nor the granting authority can be held responsible for them.

CEC acknowledges funding from Horizon Europe ERC grant no. 101125877.

VV acknowledges the support of the Research Council of Finland, Centre of Excellence in Neutron-Star Physics (grant No. 374065).

AF, GL, and GMP acknowledge support by the European Research Council (ERC) under the European Union's Horizon 2020 research and innovation programme (ERC Advanced Grant KILONOVA No. 885281).

AB and GMP acknowledge support of the Deutsche Forschungsgemeinschaft (DFG, German Research Foundation) -- Project-ID 279384907 -- SFB 1245 (subprojects B01, B06, and B07). 

GMP and OJ  acknowledge support of the Deutsche Forschungsgemeinschaft (DFG, German Research Foundation) Project-ID MA 4248/3-1.

This research was supported in part by the cluster computing resource provided by the IT Department at the GSI Helmholtzzentrum für Schwerionenforschung, Darmstadt, Germany.
We acknowledge the EuroHPC Joint Undertaking for awarding this project access to the EuroHPC supercomputer LEONARDO, hosted by CINECA (Italy) and the LEONARDO consortium, through a EuroHPC Regular Access call (proposal EHPC-REG-2026R01-209).
This work made use of the DiRAC Memory Intensive service (Cosma8) at Durham University, managed by the Institute for Computational Cosmology on behalf of the STFC DiRAC HPC Facility (www.dirac.ac.uk).
The DiRAC service at Durham was funded by BEIS, UKRI and STFC capital funding, Durham University and STFC operations grants.
DiRAC is part of the UKRI Digital Research Infrastructure.

Data analysis and plotting were done using NumPy \citep{harris_array_2020}, Polars \citep{vink_polars_2026}, Matplotlib \citep{hunter_matplotlib_2007}, and \textsc{artistools}\footnote{\url{https://github.com/artis-mcrt/artistools}} \citep{Shingles_artistools_2023}.
The manuscript was revised with suggestions from Claude large language models, but all text in this work has been approved by the authors.

\section*{Data Availability}
The ARTIS source code is publicly available \citep{artis_collaboration_artis}.
The ARTIS output files are too large to include with the arXiv source and are available from the corresponding author upon reasonable request.

\bibliographystyle{mnras}
\bibliography{references_artis,references}

\label{lastpage}
\bsp
\end{document}